\documentstyle[11pt,aaspp4]{article}
\slugcomment{To appear in Astron. J.}
\lefthead{Haisch et al.}
\righthead{A Near-Infrared L Band Survey of the Young Embedded Cluster NGC 2024}

\begin{document}

\title{A Near-Infrared L Band Survey of the Young Embedded Cluster NGC 2024}

\author{Karl E. Haisch Jr. \altaffilmark{1,2,3} and Elizabeth A. Lada 
\altaffilmark{2,4,5}}
\affil{Dept. of Astronomy, University of Florida, 211 SSRB, Gainesville, FL  
32611}

\and

\author{Charles J. Lada \altaffilmark{2,4}}
\affil{Smithsonian Astrophysical Observatory, 60 Garden Street, Cambridge, 
Massachusetts 02138}

\altaffiltext{1}{NASA Florida Space Grant Fellow}

\altaffiltext{2}{Visiting Astronomer, Fred Lawrence Whipple Observatory.}

\altaffiltext{3}{Visiting Astronomer at the Infrared Telescope Facility which is 
operated 
by 
the University of Hawaii under contract to the National Aeronautics and Space 
Administration.}

\altaffiltext{4}{Visiting Astronomer, Kitt Peak National Observatory, part of 
the National Optical Astronomy Observatories, which is operated by the 
Association of Universities for Research in Astronomy, Inc., under contract with 
the National Science Foundation.}

\altaffiltext{5}{Presidential Early Career Award for Scientists and Engineers Recipient.}

\begin{abstract}
We present the results of the first sensitive {\it L} band (3.4 $\mu$m) imaging study of the nearby young embedded cluster NGC 2024. Two separate surveys of the cluster were acquired in order to obtain a census of the circumstellar disk fraction in the cluster. We detected 257 sources to the m$_{L}$ $\leq$ 12.0 completeness limit of our $\sim$110 arcmin$^{2}$ primary survey region. An additional 26 sources with 12.0 $<$ {\it L} $<$ 14.0 were detected in the deeper survey of the central $\sim$6.25 arcmin$^{2}$ region of the cluster. From an analysis of the {\it JHKL} colors of all sources in our largest area, we find an infrared excess fraction of $\geq$ 86\%$\pm$8\%. The {\it JHKL} colors suggest that the infrared excesses arise in circumstellar disks, indicating that the majority of the sources which formed in the NGC 2024 cluster are currently surrounded by, and likely formed with circumstellar disks. The excess fractions remain very high, within the errors, even at the faintest {\it L} magnitudes from our deeper surveys suggesting that disks form around the majority of the stars in very young clusters such as NGC 2024 {\it independent} of mass. From comparison with published {\it JHKL} observations of Taurus, we find the {\it K} -- {\it L} excess fraction in NGC 2024 to be formally higher than in Taurus, although both fractions are quite high. Thus, existing {\it L} band observations are consistent with a high initial incidence of circumstellar disks in both NGC 2024 and Taurus. Because NGC 2024 represents a region of much higher stellar density than Taurus, this suggests that disks may form around most of the YSOs in star forming regions {\it independent} of environment. We find a relatively constant {\it JHKL} excess fraction with increasing cluster radius, indicating that the disk fraction is independent of location in the cluster. In contrast, the {\it JHK} excess fraction increases rapidly toward the central region of the cluster. The most likely cause for this increase is the contamination of the {\it K} band measurements by bright nebulosity in the central regions of the cluster. This suggests that caution must be applied using only {\it JHK} band observations to infer disk fractions in nebulous environments. Finally, we identify 45 candidate protostellar sources in the central regions of the cluster, and we find a lower limit on the protostellar phase of early stellar evolution in the NGC 2024 cluster of 0.4 - 1.4 $\times$ 10$^{5}$ yr, similar to that in Taurus.
\end{abstract}

\keywords{infrared: stars --- open clusters and associations: individual (NGC 
2024) --- stars: formation}

\section{Introduction}

Circumstellar disks are generally thought to play an important role in the formation and evolution of stars. In addition, disks are of particular interest since it is during the circumstellar disk phase of early stellar evolution that any subsequent planet formation will occur. In recent years, observations of star forming regions have shown that many young stellar objects (YSOs) form with disks of gas and dust (\cite{ruc85}; \cite{als87} (hereafter ALS87); \cite{bec90}; \cite{bec93}; \cite{mcs94}; \cite{kh95}; \cite{lal96}). It is therefore important to determine the frequency, nature and origin of these systems.

Young embedded clusters are likely to be responsible for producing the majority of stars in the Galaxy (\cite{lada92}, \cite{css95}, \cite{pl97}, \cite{ml99}). These young clusters contain statistically significant numbers (hundreds) of stars covering a large range in mass with similar chemical compositions in a relatively small volume at the same distance from the Sun. Therefore, clusters provide unique opportunities for the study of the early phases of star and planet formation in a statistically meaningful way. The dense stellar environments and stellar contents (ie. the presence of O stars) in clusters may affect the properties of circumstellar disks which are commonly thought to accompany the newly formed YSOs (e.g. \cite{jhb98}; \cite{sh99}). Since these disks represent the sites of any subsequent planet formation, determining the properties of disks in cluster environments is essential to evaluating the overall likelihood of planetary formation in the Galaxy.

Near-IR {\it JHK} (1.25, 1.65, 2.2 $\mu$m) color-color diagrams (i.e. a plot of 
{\it J-H} vs. {\it H-K}) have long been used as a tool for investigating the 
physical natures of YSOs in young clusters (\cite{la92}; \cite{lyg93}, \cite{ll95}, \cite{lal96}). Objects lying in the infrared excess region of these diagrams are considered to be sources with candidate circumstellar disks (\cite{la92}, \cite{kh95}, \cite{mch97}). However, {\it JHK} observations alone are not long enough in wavelength to enable complete and unambiguous disk identifications. This is due, in part, to problems arising from contamination by extended emission in HII regions, reflection nebulosity, stellar photospheric emission and source crowding in high density regions. Such effects could lead to artificially high or low disk fractions as inferred from infrared excesses in {\it JHK} color-color diagrams. Furthermore, the magnitude of the near-IR excess from a disk also depends on the parameters of the star/disk system (e.g. stellar mass/age, disk inclination, accretion rate, inner disk hole size) (ALS87; \cite{mch97}; \cite{hil98}). 

The magnitude of the infrared excess produced by a circumstellar disk increases rapidly with wavelength. Current instrument sensitivities in the {\it L} (3.4 $\mu$m) band are sufficient to detect relatively faint low mass stars due to the fact that the {\it L} band wavelength is sufficiently near the peak of the stellar energy distribution of a typical late type star. In this respect, {\it L} band observations are far superior to mid-infrared observations (i.e. $\geq$ 10 $\mu$m) made at the world's best sites with the largest telescopes. In addition, one can survey large areas in many clusters at {\it L} in reasonable amounts of telescope time, a capability not presently possible using mid-infrared detectors on ground-based telescopes. Hence, statistically significant numbers of stars in young clusters can be obtained from observations in the {\it L} band. Therefore, combining near-IR {\it JHK} observations with {\it L} band data is an extremely powerful method for investigating the nature of young stellar sources. 

In order to exploit the capabilities of {\it L} band observations to study the frequency and lifetime of circumstellar disks, we have initiated an extensive program of {\it L} band imaging surveys of nearby young clusters. In this paper, we present the results of the first sensitive {\it L} band imaging survey of the embedded cluster NGC 2024. The NGC 2024 cluster, having both a high stellar density ($\rho$$\simeq$400 stars pc$^{-3}$; \cite{lada99}) and a young mean age ($\sim$0.5 Myr; \cite{mey96}), represents an ideal region in which to study circumstellar disk properties. The NGC 2024 cluster and surrounding HII region, located in the L1630 (Orion B) giant molecular cloud at a distance of $\sim$415 pc (\cite{at82}) has been well studied. Barnes et al. (1989) have described the structure of the HII region and surrounding molecular material in detail, and their {\it K}-band (2.2 $\mu$m) imaging also revealed a cluster consisting of $\sim$30 infrared sources. A detailed near-IR {\it JHK} imaging survey conducted by Lada et al. (1991) to study global star formation in the L1630 molecular cloud yielded a large number ($\sim$300) of faint infrared sources in the NGC 2024 cluster. Subsequent {\it JHK} imaging of the central region of NGC 2024 has led to the conclusion that a large fraction, perhaps as high as two-thirds, of the low mass objects in this young embedded cluster display infrared excesses characteristic of circumstellar disks (e.g. \cite{crr96} (hereafter CRR96)).

We conducted the {\it L} band imaging survey of NGC 2024 reported here to extend the existing infrared observations to longer wavelengths in order to obtain a more complete and relatively unambiguous census of the circumstellar disk fraction in this young cluster. We present the observations and data reduction in $\S$2. In $\S$3, we discuss our methods of analysis and in $\S$$\S$4 and 5 we present the results of our imaging survey. We summarize our primary results in $\S$6.

\section{Observations and Data Reduction}

Three different sets of infrared observations were obtained for this survey. A 
summary of all observations is presented in Table \ref{sumtable}. The primary {\it L} band (3.4 $\mu$m) survey of the NGC 2024 cluster was obtained with STELIRCAM on the Fred Lawrence Whipple Observatory (FLWO) 1.2 meter telescope on Mt. Hopkins, Arizona. These data were supplemented by a deeper {\it L} band survey of the central cluster region using NSFCAM on the NASA Infrared Telescope Facility (IRTF) 3.0 meter telescope on Mauna Kea, Hawaii. The {\it JHK} (1.25, 1.65, and 2.2 $\mu$m) observations were taken with the Simultaneous Quad Infrared Imaging Device (SQIID) at Kitt Peak National Observatory on the 1.3 meter telescope. 

\subsection{STELIRCAM-FLWO-Primary {\it L} Band Survey}

Observations of the NGC 2024 cluster were made with STELIRCAM during the periods 1998 January 06-08 and 1998 December 03-08 respectively. STELIRCAM consists of 
two 256$\times$256 InSb detector arrays. Each is fed from a dichroic that separates 
wavelengths longer and shorter than 1.9 $\mu$m. Three separate magnifications 
can be selected by rotating different cold lens assemblies into the beam. For 
our primary survey we selected a field of view of 2.5\arcmin$\times$2.5\arcmin \hspace*{0.05in}with a resolution of 0.6\arcsec \hspace*{0.05in}per pixel.

Twenty five fields arranged in a 5$\times$5 square, covering an area of $\sim$110 arcmin$^{2}$ were observed toward the cluster. The individual images were overlapped 
by 30\arcsec \hspace*{0.05in}in both Right Ascension and Declination. This overlapping provided redundancy for the photometric measurements of sources located in the overlapped regions and enabled an accurate positional 
mosaic of the cluster to be constructed. Each individual field in the mosaic was observed in a 3$\times$3 square pattern with 12\arcsec \hspace*{0.05in}offsets or dithers between 
positions. The pattern was observed such that the telescope was not offset in 
the same direction twice. Once the 9 dithered positions had been observed, 
the telescope was offset by 5\arcsec \hspace*{0.05in}and the pattern repeated. The integration time 
at each dither position was 10 sec (0.1 x 100 coadds) with a total integration 
time for each field of 3 minutes. Four of the 25 fields, covering the inner
$\sim$20 sq. arcmin of the cluster, were observed for a total integration time of 18 minutes. These fields were observed using the method described above with the inclusion of a 7\arcsec \hspace*{0.05in}dither between each 3 minute sequence.

All data were reduced using the Image Reduction and Analysis Facility (IRAF). An average dark frame was constructed from the darks taken at the beginning and end of each night's observations. This dark frame was subtracted from all target 
observations to yield dark subtracted images. Sky frames were individually made 
for each observation by median combining the nearest nine frames in time to the 
target observation. Each sky frame was checked to confirm that all stars had 
been removed by this process. The sky frames were then normalized to produce 
flat fields for each target frame. All dark subtracted target frames were then 
processed by subtracting the appropriate sky frame and dividing by the flat 
field. Finally all target frames for a given position in the cluster were 
registered and combined to produce the final image of each field.

\subsection{NSFCAM-IRTF-Deep {\it L} Band Survey}

Additional {\it L} band observations of the cluster were made with NSFCAM during the period 1999 January 08 - 10. NSFCAM consists of a single 256$\times$256 InSb detector array. Three different magnifications can be selected by rotating 
different cold lens assemblies into the beam. For our deep study, we used a plate 
scale of 0.3\arcsec \hspace*{0.05in}per pixel with a corresponding field of view of approximately 1.25\arcmin$\times$1.25\arcmin.

Four fields centered on the cluster were observed in a 2$\times$2 square covering an 
area of $\sim$6.25 arcmin$^{2}$. The fields were observed using the same method as 
described above for the STELIRCAM observations with the exception that, between 
each dither position, the telescope was nodded to separate sky positions 300\arcsec \hspace*{0.05in}east of the cluster. The total integration time for each field was $\sim$4.6 minutes. The data were reduced in the same manner as that for STELIRCAM, however in this case, the sky frames were constructed by median combining the nearest 9 sky positions in time to the target observation. The sky frames were checked to ensure that all stars had been removed and normalized to produce flat fields. After sky subtracting and flat fielding, the 18 individual positions for each cluster field were registered and combined to produce the final image.

\subsection{SQIID-NOAO-{\it JHK} Survey}

The {\it JHK} observations for our survey were taken with SQIID in the period 
1992 January 12 - 15. SQIID consists of four 256$\times$256 PtSi detector arrays which are mounted to permit simultaneous observations at {\it J, H, K}, and {\it 
L}. Since the {\it L} band saturates quickly due to the high sky background, 
only the {\it J, H}, and {\it K} bands were used. On the 1.3 m telescope, SQIID 
provides an imaging scale of 1.36\arcsec \hspace*{0.05in}per pixel at {\it K} band with a field of view of 5.5\arcmin$\times$5.5\arcmin.

Twenty five SQIID fields were observed in a 5$\times$5 mosaic covering an area of $\sim$462 arcmin$^{2}$ on the sky. Each SQIID field was overlapped by 1.5 
arcminutes in both Right Ascension and Declination, allowing for both redundancy of photometric measurements of sources located in the overlapped regions and 
accurate positional placement of the mosaicked fields. Eighteen control fields (covering an area of $\sim$545 arcmin$^{2}$) located well away from the cluster were observed in order to account for the distribution of foreground/background field stars. These fields were found to be free of significant molecular material by examination of the Palomar Sky Survey Prints. The integration time in each filter was 3 minutes for observations of both cluster and control fields. Each field was observed twice with a 15\arcsec \hspace*{0.05in}dither between observations. Each pair of dithered images were combined to produce images with an effective integration time of 6 minutes in each filter. The data were reduced using IRAF routines as described in Lada \& Lada (1995). In this paper, we restrict our analysis of the SQIID data to the central $\sim$110 arcmin$^{2}$ area corresponding to our SAO survey region.

\section{Analysis}

\subsection{Source Extraction and Photometry}

For each of the three data sets, infrared sources were identified using the 
DAOFIND routine (\cite{stet87}) within IRAF. The full width at half maximum (FWHM) of the point spread function (PSF) for the SQIID data was typically $\sim$2.0 pixels while the FWHM of the PSF for the STELIRCAM and NSFCAM data ranged from 2.3 to 2.8 pixels. DAOFIND was run on each individual image in each dataset using a FWHM of the PSF between 2.0 and 2.5 pixels and a single pixel finding threshold equal to 3 times the mean noise of each image. Each frame was individually inspected and the DAOFIND coordinate files were edited to remove bad pixels and any objects misidentified as stars, as well as add any missed stars to the list. Aperture photometry was then performed using the routine PHOT within IRAF. We used an aperture of 4 pixels in radius for the STELIRCAM and NSFCAM photometry and a 3 pixel radius aperture for the SQIID photometry. Aperture diameters were selected such that they were at least twice the FWHM of the PSF of the stars. The sky values around each source were determined from the mode of intensities in an annulus with inner and outer radii of 10 and 20 pixels respectively. Our choice of aperture sizes ensured that the flux from at least 90\% of the stars was not contaminated by the flux from neighboring stars, however they are not large enough to include all the flux from a given source. In order to account for this missing flux, aperture photometry was performed on all bright, isolated sources using the same aperture used for the photometry of the standard stars, 10 pixels for each dataset. Fluxes in both the large and small apertures were compared and the instrumental magnitudes for all sources were corrected to account for the missing flux.

Photometric calibration of all three datasets was accomplished using the list of standard stars of Elias et al. (1982). The standards were observed on the same nights and through the same range of airmasses as the cluster and control fields. Zero points and extinction coefficients were established for each filter for a given night. The photometric system of NSFCAM is very similar to the UKIRT system. We have assumed SQIID to be on the CIT system (ie. \cite{hll97}). Therefore, no color corrections have been applied. We compared the {\it L} band magnitudes of the seventy one stars common to our STELIRCAM and NSFCAM datasets. At the completeness limit of the STELIRCAM survey, the photometry between the STELIRCAM and NSFCAM {\it L} band observations agreed to within 10\%.

\subsection{Completeness Limits}

The completeness limits of the three sets of observations were determined by adding artificial stars to each of the images and counting the number of sources recovered by DAOFIND. In regions where source crowding was not a problem, artificial stars were added at random positions to each image in twenty separate half magnitude bins with each bin containing one hundred stars. The twenty bins covered a magnitude range from 10.0 to 20.0. The artificial stars were examined to ensure that they had the same FWHM of the PSF as the real sources in the image.  Aperture photometry was performed on all sources to confirm that the assigned magnitudes of the added sources agreed with those returned by PHOT. All photometry agreed to within 0.1 magnitudes. DAOFIND and PHOT were then run and the number of identified artificial sources within each half magnitude bin was tallied. This process was repeated 20 times. In the central regions of the cluster, where source crowding was a concern, fifty sources were added at random positions to each image. The procedure for estimating the completeness limit was the same as described above, however the process was repeated 40 times. We estimate that the identification of sources in our SQIID survey, averaged over regions without nebulosity, is 90\% complete to {\it m}$_{J}$ = 16.5, {\it m}$_{H}$ = 15.5, and {\it m}$_{K}$ = 14.5. In the central 18\% of the cluster, where nebulosity is a concern, the completeness limits are 0.5 magnitudes brighter at {\it J,H}, and {\it K}. For our STELIRCAM {\it L} band data, the estimated 3 and 18 minute integration completeness limits are {\it m}$_{L}$ = 12.0 and {\it m}$_{L}$ = 12.9 respectively, while our NSFCAM observations are complete to {\it m}$_{L}$ = 14.0.

\subsection{Astrometry}

For each individual cluster source in our three separate surveys, $\alpha$ and $\delta$ positions were determined. A centering algorithm in APPHOT was used to obtain center positions for the sources in pixels. Many sources were present in the overlap regions of our NGC 2024 fields. Duplicate stars on adjacent frames were used to register the relative pixel positions of the individual frames in each survey onto a single positional grid. Pixel positions were converted to equitorial coordinates using the positions of NGC 2024 IRS1 and IRS2 as references and plate scales of 1.36\arcsec/pixel ({\it K} band), 0.6\arcsec/pixel, and 0.3\arcsec/pixel for SQIID, STELIRCAM, and NSFCAM respectively. A comparison of the source coordinates between each of the three surveys yields an agreement to within 1\arcsec. A list of all objects for which we have complete {\it JHKL} photometry is given in Table \ref{phottable}. Also listed in Table \ref{phottable} are the RA (1950) and Dec (1950) coordinates and the near-infrared colors for each source. Absolute coordinates are good to only a few arcseconds and depend on the accuracy of the IRS1 and IRS2 positions (taken from the IRAS Point Source Catalog). However, our analysis does not depend on the astrometry.

\section{Results}

\subsection{Spatial Distribution}

We detected 257 and 107 sources at {\it L} band with STELIRCAM to the completeness limits of our $\sim$110 and $\sim$20 arcmin$^{2}$ surveys ({\it L} = 12.0 and 12.9 respectively). All of these sources were detected in at least one other band (typically {\it K}). Within the same area covered by our STELIRCAM observations ($\sim$110 arcmin$^{2}$), we detected a total of 286 {\it J} band sources ({\it m$_{J}$}$<$ 16.5), 311 {\it H} band sources ({\it m$_{H}$}$<$ 15.5) and 343 {\it K} band sources ({\it m$_{K}$}$<$ 14.5) in our SQIID survey. Our NSFCAM observations of the central $\sim$6.25 arcmin$^{2}$ region of the NGC 2024 cluster reveal 26 additional sources with 12.0 $<$ {\it L} $<$ 14.0. Four of these had 12.0 $<$ {\it L} $<$ 12.9, and were thus detected in our $\sim$20 arcmin$^{2}$ STELIRCAM survey region. Of the 26 additional sources detected with NSFCAM, 20 sources were not detected to the completeness limit of our SQIID observations. However, Levine, Lada, \& Elston (2000, in preparation) have recently obtained {\it K} band tip-tilt data of the NGC 2024 cluster using the 4m telescope at Cerro Tololo Interamerican Observatory (CTIO). This dataset extends to a {\it K} band completeness limit of {\it K} $\simeq$ 16.5. From these observations, we can identify {\it K} band counterparts for all but one of the NSFCAM {\it L} band sources.

Figures \ref{N2024_L}(a) and \ref{N2024_L}(b) (Plate \#\#) present an {\it L} band mosaic and a {\it JHK} color mosaic of the NGC 2024 cluster. These images are displayed with north at the top and west on the right. The spatial distribution of all sources to {\it K} = 14.5 in our SQIID/STELIRCAM survey is shown in Figure \ref{clusposn}. Sources which are detected at both {\it K} and {\it L} are plotted using a point while those sources which are visible only at {\it K} in our SQIID data are shown with a star. A slight central condensation of sources may be present, otherwise the distribution is fairly uniform with no obvious differences in the locations of sources detected and not detected at {\it L}. In Figure \ref{cluslonly} we plot the inner 2\arcmin x 2\arcmin \hspace*{0.05in}region of the cluster observed in our NSFCAM survey. All sources which were detected to the SQIID completeness limit in the central frame ({\it K} = 14.0) are shown by a point. Sources having {\it K} magnitudes fainter than {\it K} = 14.0 (obtained from Levine, Lada, \& Elston (2000, in preparation) are shown by a triangle. A star denotes the position of the source detected only at {\it L} band. 

\subsection{Color-Color Diagrams}

In Figures \ref{ccjhkl}(a) and \ref{ccjhkl}(b) we present the {\it JHK} and {\it JHKL} color-color diagrams for the NGC 2024 cluster. In both diagrams, we have included only those 142 sources from our SQIID survey which have {\it K} magnitudes equal to or brighter than the completeness limit of our STELIRCAM {\it L} band survey (ie. {\it K} $\leq$ 12.0). Examining only those sources for which we are sensitive to stellar photospheres (ie. {\it K} -- {\it L} = 0.0), allows meaningful comparisons between the color-color diagrams to be made for the purpose of obtaining the fraction of the sources having an infrared excess, and thus the circumstellar disk fractions (see Discussion). Figure \ref{ccjhk} presents the {\it JHK} color-color diagram for the 77 control field sources which match the above criteria. In these figures, we plot the locus of points corresponding to the unreddened main sequence as a solid line and the locus of positions of giant stars as a heavy dashed line (Bessell \& Brett 1988). The two leftmost parallel dashed lines define the reddening band for main sequence stars and are parallel to the reddening vector. Crosses are placed along these lines at intervals coresponding to 5 mag of visual extinction. The classical T Tauri star (CTTS) locus is plotted as a dot-dashed line (\cite{mch97}). The reddening law of Cohen et al. (1981), derived in the CIT system and having slopes in the {\it JHK} and {\it JHKL} color-color diagrams of 1.692 and 2.750 respectively, has been adopted.

Clear differences exist between the {\it JHK} cluster and control field diagrams. The stars in the {\it JHK} control fields are tightly concentrated near the unreddened main sequence with spectral types corresponding to late G to late M stars. A smaller portion of stars are found around the giant branch. This indicates that the control fields consist of field stars which possess very little interstellar reddening. In contrast, the stars in the NGC 2024 cluster are displaced from the main sequence and spread over a larger area of the reddening band. This displacement is due to extinction by the molecular cloud associated with NGC 2024. Comparison of the mean {\it H} -- {\it K} colors of the cluster and control field stars yields an average extinction of A$_{V}$ = 10.4 $\pm$ 1.9 for all sources in the reddening band of the {\it JHK} color-color diagram. Furthermore, the spread of the cluster stars within the reddening band indicates that there is differential extinction with a range corresponding to roughly 0 - 30 visual magnitudes. This range in extinction toward the cluster, combined with the fact that statistically the majority of the stars observed in the cluster field are actual cluster members (as determined from a {\it K} band luminosity function study of the NGC 2024 cluster by Dahari \& Lada (2000, in preparation), suggests that our estimate of the extinction to the background stars is likely an underestimate and that the cluster stars themselves are typically extincted by A$_{V}$$\simeq$10.5.

A significant fraction of the cluster sources fall outside and to the right of 
the reddening lines in the infrared excess region of the color-color diagrams. In the {\it JHKL} diagram, a total of 112/142 (79\%$\pm$8\%) of the sources lie in the infrared excess region.  Inspection of the {\it JHK} color-color diagram for the cluster reveals a smaller fraction of infrared excess sources. Indeed, 76/142 (54\%$\pm$6\%) of the sources lie in the infrared excess region of the {\it JHK} diagram. 

We must correct these fractions for field star contamination. Since we are only considering those stars to {\it K} = {\it L} = 12, the correction is the same for the {\it JHK} and {\it JHKL} fractions. The foreground/background contamination was determined by first counting the total number of stars observed in the control fields to {\it K} = 12. Scaling this number to the cluster area observed and accounting for the average extinction in NGC 2024 (A$_{K}$$\simeq$1.0 magnitudes), we find a field star contamination of eleven stars. Applying this correction to the infrared excess fractions, we estimate that 86\%$\pm$8\% and 58\%$\pm$7\% of the sources in the NGC 2024 cluster have colors suggestive of {\it JHKL} and {\it JHK} infrared excess emission respectively. A similar large difference in the {\it JHKL} and {\it JHK} excess fractions (84\%$\pm$10\% versus 64\%$\pm$9\%) is observed in our deeper ({\it K} = {\it L} = 12.9) STELIRCAM data. On the other hand, the {\it JHKL} and {\it JHK} excess fractions are the same within the errors (90\%$\pm$15\% and 87\%$\pm$15\%) for our NSFCAM survey which covers a much smaller region and extends to {\it K} = {\it L} = 14.0 (the completeness limit in the central region of the cluster). A list of the {\it JHK} and {\it JHKL} infrared excess fractions for the three areas and completeness limits of our survey is given in Table \ref{fractable}.

The derived fraction of excess sources is sensitive to the location of the boundary of the reddening band. This depends on the adopted intrinsic colors of the latest spectral type stars. The dwarf main sequence colors used by different authors become disparate at the latest spectral types, starting at about M3 (e.g. \cite{kor83}; \cite{bb88}). By a spectral type of M5, a comparison of the {\it H} -- {\it K} and {\it K} -- {\it L} colors from Koornneef (1983) and Bessell \& Brett (1988) yields differences of 0.03 and 0.11 magnitudes respectively. The excess fractions derived above were calculated assuming the boundary of the reddening band in the {\it JHKL} and {\it JHK} diagrams pass through M5 and A0 colors from Bessell \& Brett (1988) respectively. A0 rather than M5 colors were used in the {\it JHK} diagram since if one extends the righthand reddening line from an M5 spectral class toward bluer colors, some of the earlier type main sequence stars fall in the infrared excess region of the diagram. This reddening line extends through main sequence colors which are somewhat later than M5. However, we consider the {\it JHKL} fraction as derived above to be a lower limit on the true excess fraction. The average extinction in the cluster is A$_{K}\simeq$1.0, and therefore the unextincted magnitudes of the sources at our completeness limit are in fact {\it K} = 11.0 rather than {\it K} = 12.0. At the distance and age of NGC 2024 ($\sim$1 Myr), this implies spectral types of M0--M3 from {\it K} band luminosity function models (A. Muench, private communication). This is consistent with the near-infrared spectral types from Meyer (1996) for 33 sources having {\it K} $\leq$ 12.0. All sources had spectral classes earlier than M4 (including errors in spectral type). If we assume M4 colors for the right-hand reddening vector, the {\it JHKL} fraction increases to 92$\pm$8\%.

The magnitude and frequency of near-IR excess sources determined from color-color diagrams also depends on the adopted reddening law and to a lesser extent on the photometric system used to plot the positions of the main sequence stars and the reddening bands. Our adopted reddening law of Cohen et al. (1981) is strictly true for the CIT photometric system, since this is the system in which this reddening law was derived. However, if the reddening law for the NGC 2024 cluster were 10\% less steep, similar to that for the $\rho$ Oph region (ie. E$_{J-H}$/E$_{H-K}$ = 1.57; \cite{klb98}), we find {\it JHK} and {\it JHKL} excess fractions of 47$\pm$6\% and 82$\pm$8\% respectively. Within the quoted (statistical) errors, these fractions are the same as those we derived using a Cohen et al. (1981) reddening law. In summary, although the {\it absolute} excess fractions derived for a given cluster will depend somewhat on the choice of the boundary, comparisons of the {\it apparent} excess fractions with other clusters will be valid as long as we are consistent in selecting reddening laws and main sequence colors, and as long as the reddening law is the same for the clusters being compared. This latter condition is likely to hold since at infrared wavelengths, the extinction law is generally not observed to vary from place to place in the Galaxy (\cite{mat90}).

\subsection{Sources with Anomalous {\it K} -- {\it L} Colors}

In the {\it JHKL} and {\it JHK} color-color diagrams, approximately 4 sources (3\% of the sample) have colors which place them in the ``forbidden'' regions either to the left of the main sequence reddening band or to the right and below the CTTS locus. Typical causes for the presence of stars in these forbidden regions are contamination of the photometry due to unresolved binaries and large photometric errors. Two of the four anomalous sources were found to be blended stars which are unresolved in our images. The other two sources were found to be single stars with small photometric errors. Thus, large photometric errors and unresolved binaries can account for half of the anomalous sources.

Since the {\it L} band observations were taken roughly seven years after the {\it JHK} data, source variability could also cause a star's placement in one of the forbidden regions. To test this possibility, we compared the {\it JHK} magnitudes of the 40 sources common to both our survey and that of \cite{mey96}. Less than 10\% were found to have magnitudes which differed by more than 0.2 magnitudes in all three bands. This can account for all the sources with anomalous colors in our {\it JHKL} color-color diagram. In addition, variability would cause an equal number of sources to be pushed into the forbidden and excess regions of the color-color diagram. However, the vast majority of the sources in NGC 2024 exhibit infrared excesses. Finally, variability has also been shown to have a negligible effect on the infrared excesses found in other star forming regions (e.g. \cite{bar97}, \cite{her00}, Lada et al. 2000, in preparation). Therefore, we find it very unlikely that source variability is a major factor in producing infrared excesses.

\subsection{Infrared Excess Fraction as a Function of Cluster Radius}

In Figure \ref{projrad} we present the infrared excess fractions determined from both {\it JHKL} and {\it JHK} color-color diagrams as a function of projected cluster radius as determined from our {\it K} = {\it L} = 12.0 completeness limit data in NGC 2024. The diagram was constructed by calculating the excess fractions in successively larger areas, increasing the projected radius by 0.1 pc each time. The excess fractions have been corrected for field star contamination. The dot-dashed line indicates the excess fraction as determined from the {\it JHKL} colors while the dashed line represents the excess fraction determined from the {\it JHK} colors. The error bars indicate $\sqrt{{\it N}}$ errors in the excess fractions. A decrease in the infrared excess fraction with projected cluster radius as obtained from the {\it JHKL} colors is observed, however it is not significant to within the errors. The decrease observed in the {\it JHK} excess fractions {\it is} significant. The excess fraction declines out to a radius of $\sim$0.4 pc, at which point the fractions remain constant. These trends in the infrared excess fractions with projected cluster radius also hold if we consider the case in which the extinction in the outer regions of the cluster is lower than that near the center (e.g. A$_{V}$ = 5 mag rather than the average for the cluster of A$_{V}$ $\sim$ 10 mag); that is, if the correction for background star contamination were larger in the outer regions of the cluster than in the inner regions.

\subsection{Distribution of {\it K} -- {\it L} Colors}

In Figure \ref{ngc2024coldist} we present the {\it K} -- {\it L} frequency distribution for the 185 NGC 2024 cluster sources from our combined STELIRCAM and NSFCAM surveys for which {\it K} -- {\it L} colors could be determined. The shaded region in the distribution represents those sources in our NSFCAM survey for which {\it K} magnitudes were obtained from Levine, Lada, \& Elston (2000, in preparation) (ie. 14.0 $<$ {\it K} $<$ 16.5). These objects represent the reddest sources in the survey. A broad peak, centered around {\it K} -- {\it L} $\simeq$ 0.8 - 1.0 appears in the distribution, with only a small fraction of the sources having {\it K} -- {\it L} $\geq$ 2.0. In our STELIRCAM survey, the number of sources having {\it K} -- {\it L} $\geq$ 0.5, 1.0, 1.5 and 2.0 is 147, 82, 27 and 4 respectively. All of the sources in our NSFCAM survey have {\it K} -- {\it L} $\geq$ 1.0, 19 have {\it K} -- {\it L} $\geq$ 1.5 and 10 have {\it K} -- {\it L} $\geq$ 2.0. Four of the sources have {\it K} -- {\it L} $\geq$ 4.0, including the one source which was only detected at {\it L}. 

\subsection{Is Our Cluster Sample Representative?}

Are we seeing a representative sample of the stars to {\it K} = {\it L} = 12.0? Put another way, are we detecting all of the stars present in the cluster to our completeness limit, or are we missing a large fraction of the sources because they are embedded deeper in the molecular cloud and therefore extincted beyond our detection limit? In order to investigate this question, we examine the {\it K} band magnitude distribution of the sources in the large clump between A$_{V}$ = 5 -- 15 mag in the reddening band of the {\it JHK} color-color diagram for the cluster presented in Figure 4(a). We count the number of sources we expect to observe to {\it K} = 12.0 after including an additional amount of extinction equal to the average extinction found for the cluster (A$_{V}$ $\simeq$ 10.5; see previous section). We compare this predicted number of sources to the actual number observed in the reddening band of the color-color diagram between A$_{V}$ = 15 -- 25 mag. For the NGC 2024 cluster, we find a total of 23 stars with {\it K} $\leq$ 12.0 after artifically extincting the {\it J,H} and {\it K} magnitude distributions. An examination of the color-color diagram reveals that only 7 stars are actually observed in the range of visual extinctions between 15 -- 25 mag. A similar predicted vs. observed result holds true even if we consider all sources to the completeness limit of the SQIID survey ({\it K} = 14.5). Thus, we predict that we should detect a larger number of sources to our completeness limit if the NGC 2024 cluster were more embedded in the molecular cloud than we actually observe. We therefore conclude that we are seeing a representative sample of the population of stars to {\it K} = 12.0 in the NGC 2024 cluster. 
\clearpage

\section{Discussion}

\subsection{Infrared Excess Fractions: A Circumstellar Disk Census}

The infrared excess fraction determined from the {\it JHKL} color-color diagram for the entire area surveyed in NGC 2024 ($\geq$ 86\%$\pm$8\%) is much larger than the 58\%$\pm$7\% found from the {\it JHK} diagram. Predictions from both observations and modelling suggest that this is what one would expect from excess emission from circumstellar disks (\cite{la92}, \cite{mch97}).

Kenyon \& Hartmann (1995) have compiled observations of a sample of $\sim$170 YSOs in the Taurus-Auriga dark cloud, which include the spectral energy distribution (SED) classification for each source as either Class I (YSOs surrounded by massive circumstellar envelopes of infalling material), Class II (YSOs surrounded by optically thick circumstellar disks) or Class III (diskless sources with no infrared excess emission)(ALS87). In Figures \ref{tauclassjhk}(a) and \ref{tauclassjhk}(b), we present the {\it JHK} and {\it JHKL} color-color diagrams for those 136 sources with {\it m}$_{K}$ $\leq$ 9.5, the completeness limit of our NGC 2024 sample corrected for the distance to Taurus (d = 140pc). Only 59/86 (69\%$\pm$9\%) of the Class II sources lie in the infrared excess region of the {\it JHK} color-color diagram, whereas {\it all} of the Class II sources have colors indicative of excess emission in the {\it JHKL} diagram. Therefore, the {\it JHKL} colors are more robust for identifying circumstellar disks than {\it JHKL} colors.

The measured overall {\it JHK} and {\it JHKL} excess fractions in Taurus are 49\%$\pm$6\% and 69\%$\pm$7\%. Thus, in Taurus, the {\it JHKL} excess fraction for all sources with {\it m}$_{K}$ $\leq$ 9.5 is higher than the {\it JHK} fraction, similar to our observations of NGC 2024. This closely resembles the expected behavior of infrared excess emission produced by a population of sources characterized by a high disk fraction. Therefore, these measurements strengthen the suggestion that the infrared excesses we observe in NGC 2024 originate in circumstellar disks. Moreover, the disk fraction derived from our {\it JHKL} observations is a relatively accurate representation of the disk fraction in the NGC 2024 cluster.

The large {\it JHKL} excess fraction ($\geq$86\%) in NGC 2024 suggests that a majority of the sources which formed in the cluster were initially surrounded by disks, and remain surrounded by disks to the present time. Such a large disk fraction argues for a very young age for the NGC 2024 cluster, consistent with the young cluster age (mean age $\simeq$0.3 Myr) obtained by Meyer (1996). Furthermore, a very large {\it JHKL} excess fraction ($\geq$84\%) is observed even if we consider our deeper surveys (c.f. Table \ref{sumtable}). Recall that in NGC 2024, we are likely detecting sources to spectral types of at least M3, consistent with the observations of Meyer (1996) for the sources common to our survey and theirs. The fact that the infrared excess fraction remains very high to within the errors at the faintest {\it L} magnitudes suggests that disks form around the majority of the stars in very young clusters such as NGC 2024 {\it independent} of mass. 

The excess fraction measured for NGC 2024 is formally higher than that in Taurus. Nevertheless, both excess fractions are quite high. The NGC 2024 cluster represents an environment with an order of magnitude higher stellar density relative to Taurus, where star formation occurs in relative isolation (\cite{lada99}). This suggests that a high stellar density does not appear to inhibit the formation of circumstellar disks.

\subsection{Disk Fraction vs. Projected Cluster Radius}

In Figure \ref{projrad}, we see that the {\it JHKL} infrared excess fraction is relatively constant, within the errors, as a function of projected cluster radius while there is a significant decrease in the {\it JHK} excess fraction. Since the {\it K} -- {\it L} color is apparently able to identify {\it all} sources with candidate circumstellar disks in a {\it JHKL} color-color diagram, we can determine when the disk fraction begins to decrease. The fact that the {\it JHKL} excess fraction remains constant with projected radius implies that the disk fraction does not decrease to the boundary of our survey. The observed trend in the {\it JHK} excess fraction, however, appears to depend on projected cluster radius, perhaps suggesting that the efficiency of using {\it JHK} observations as a diagnostic in identifying disks varies with radius.

Lada \& Adams (1992) and Meyer, Calvet, \& Hillenbrand (1997) have shown that disk orientation and accretion rate affects the observed {\it JHK} colors, and hence whether or not a source is identified as having an infrared excess. The disk orientation and accretion rates could be different in the cluster center where the stellar density is higher than in the outer regions. In particular, close encounters between stars in the central region of the cluster may lead to tidal effects which increase the disk accretion rates and thus the disk detectability (\cite{ost94}). The NGC 2024 cluster has a high central density of $\sim$6$\times$10$^{3}$ M$_{\odot}$ pc$^{-3}$, thus the effects of induced accretion may be a contributing factor to the high apparent {\it JHK} disk fraction observed in the central regions of the cluster. 

However, the NGC 2024 cluster contains large amounts of nebulosity, especially near the cluster center, and this may be responsible for the observed trend in the {\it K} band excess fraction with cluster radius, rather than an increase in {\it K} band efficiency in detecting disks in the central cluster region. To test this, we added 300 artificial stars with {\it JHKL} magnitudes from 9.0 to 14.0 and having main sequence colors to regions of the cluster with and without nebulosity. Aperture photometry was performed on these sources and their colors were calculated. Approximately 23\% of the sources, independent of magnitude, which were placed in regions with nebulosity exhibited colors indicative of infrared excesses in the {\it JHK} color-color diagram, while all sources placed in non-nebulous regions showed no excesses. Sources in non-nebulous regions also show no excesses in the {\it JHKL} diagram. In nebulous regions, however, only sources with {\it L} $>$ 11.0 exhibited {\it JHKL} excesses ($\sim$15\% excess fraction). Therefore, the large {\it JHK} excess fraction observed in the cluster center relative to the outer regions may be due to contamination caused by the extended nebulosity. The {\it K} -- {\it L} color is less susceptible to this effect and hence better represents the observed trend in disk fraction with cluster radius.

Hillenbrand et al. (1998) have also observed an increase in the disk frequency toward the center of the Orion Nebula Cluster (ONC) using a diagnostic based on {\it I$_{C}$} -- {\it K} colors, which they attribute to the effects of the high stellar density near the cluster center (the mean distance between stars is only $\sim$1000 AU). However, it is also possible that the large amount of nebulosity in the ONC may be producing artificial infrared excesses similar to that observed in the NGC 2024 cluster, and hence a larger {\it apparent} disk fraction near the center of the ONC. Therefore, we suggest that caution must be applied using only observations shorter than {\it K} band to infer disk fractions in nebulous environments.

\subsection{Nature of Very Red Sources}

Thirteen sources in our STELIRCAM survey have {\it K} -- {\it L} colors $\geq$ 1.5 and lie in the region of the {\it JHKL} color-color diagram beyond the reddening line projected from the red edge of the CTTS locus (\cite{mch97}). An additional 14 very red sources also have {\it K} -- {\it L} $\geq$1.5 and are located for the most part high within the CTTS reddening band. Furthermore, 18 sources from our NSFCAM survey have similar {\it K} -- {\it L} colors, however {\it J} -- {\it H} colors have yet to be determined for these sources and hence they cannot be plotted on a color-color diagram. Kenyon \& Hartmann (1995) have determined that all of the sources in Taurus-Auriga having {\it K} -- {\it L} $\geq$ 1.5 were Class I sources. If the sources in NGC 2024 are similar to those in Taurus-Auriga, it is possible that most, if not all, of these 45 sources could be Class I (protostellar) objects.

If the star formation rate in NGC 2024 has been constant, and the age of a typical T Tauri star (TTS) is known, then an estimate of the Class I (protostellar) phase of pre-main sequence evolution can be obtained. The Class I lifetime is given by: 

\begin{equation}
\tau_{Class I} = \frac {N_{Class I}}{N_{PMS}} \times \tau_{Cluster} \hspace*{0.1in}yr.
\end{equation}

There are 45 candidate Class I sources out of 328 PMS stars having {\it K}$\leq$14.0 identified in the NGC 2024 cluster. If TTS ages are $\sim$0.3 - 1 Myr (Meyer 1996), then the Class I lifetime is $\tau_{Class I}$ $\sim$ 0.4 - 1.4 $\times$ 10$^{5}$ yr. Our estimate of the Class I lifetime is a lower limit since the number of candidate Class I sources was derived from our combined STELIRCAM/IRTF sample which constitutes two regions with differing completeness limits. Thus, a larger number of Class I sources may be present in the cluster. Nevertheless, our estimate is similar to the approximate Class I lifetime of 1 - 2 $\times$ 10$^{5}$ yr in Taurus-Auriga (e.g. \cite{myer87}; \cite{ken90}; \cite{ken94b}; \cite{kh95}).

\section{Conclusions}

We have conducted the first systematic {\it L} band survey of a large area of the young embedded cluster NGC 2024 in order to obtain a relatively unambiguous census of the circumstellar disk fraction in the cluster. Our main results can be summarized as follows:

1. We detect a total of 257 and 107 {\it L} band sources to the {\it L} = 12.0 and 12.9 completeness limits of our $\sim$110 and $\sim$20 arcmin$^{2}$ STELIRCAM surveys. In our deeper NSFCAM survey, we find 26 additional sources with 12.0$<${\it L}$<$14.0 in the central 6.25 arcmin$^{2}$ region of the cluster.

2. An analysis of the {\it JHK} and {\it JHKL} color-color diagrams for the entire area surveyed in the NGC 2024 cluster reveals infrared excess fractions of 58\%$\pm$7\% and 86\%$\pm$8\% respectively. This closely resembles the behavior of infrared excess emission produced by circumstellar disks. Thus, this suggests that the infrared excesses in NGC 2024 originate in such disks, and that the combined {\it JHKL} observations provide a more robust estimate of the circumstellar disk fraction in NGC 2024 than {\it JHK} colors alone.

3. Within the errors, our data suggests that the majority of the sources which formed in the NGC 2024 cluster were surrounded by circumstellar disks, and these disks are still present around these sources. In addition, very large {\it JHKL} excess fractions ($\geq$84\%) are also found when we consider our deeper surveys. Given that the excess fractions remain very high to within the errors at the faintest {\it L} magnitudes, we suggest that disks formed around the majority of the stars in NGC 2024 {\it independent} of mass.

4. The {\it K} -- {\it L} excess fraction measured for NGC 2024 is formally higher than that in Taurus, although both fractions are quite large. The NGC 2024 cluster represents an environment with a higher stellar density relative to Taurus, where star formation occurs in relative isolation. This suggests that disks form around most of the YSOs in star forming regions {\it independent} of environment.

5. A decrease in the {\it JHKL} circumstellar disk fraction with projected cluster radius is observed, however it is not significant to within the errors. This suggests that the disk fraction does not decrease to the boundary of our survey. In contrast, the {\it JHK} excess fraction increases rapidly toward the central region of the cluster. The most likely cause for this increase is the contamination of the {\it K} band measurements by bright nebulosity in the central regions of the cluster. Observations at {\it L} are not as susceptible to this effect and hence better represent the observed trend in disk fraction with cluster radius. This suggests that caution must be applied using only {\it JHK} band observations to infer disk fractions in nebulous environments.

6. Forty five sources from our combined {\it L} band STELIRCAM/IRTF survey have large {\it K} -- {\it L} colors characteristic of possible Class I sources. If all of these objects are Class I sources, we calculate the Class I lifetime to be $\sim$0.4 - 1.4 $\times$10$^{5}$ yr. This is similar to the approximate Class I lifetime of 1 - 2 $\times$ 10$^{5}$ yr in Taurus-Auriga.

\acknowledgements

We would like to thank Joanna Levine for generously providing infrared photometry from CTIO in advance of publication. We also thank Eric Tollestrup for the development, use and support of STELIRCAM. K. E. H. gratefully acknowledges support from a NASA Florida Space Grant Fellowship and an ISO grant through JPL \#961604. E. A. L. acknowledges support from a Research Corporation Innovation Award and a Presidential Early Career Award for Scientists and Engineers (NSF AST 9733367) to the University of Florida. We also acknowledge support from and ADP (WIRE) grant NAG 5-6751.

\newpage
\figcaption[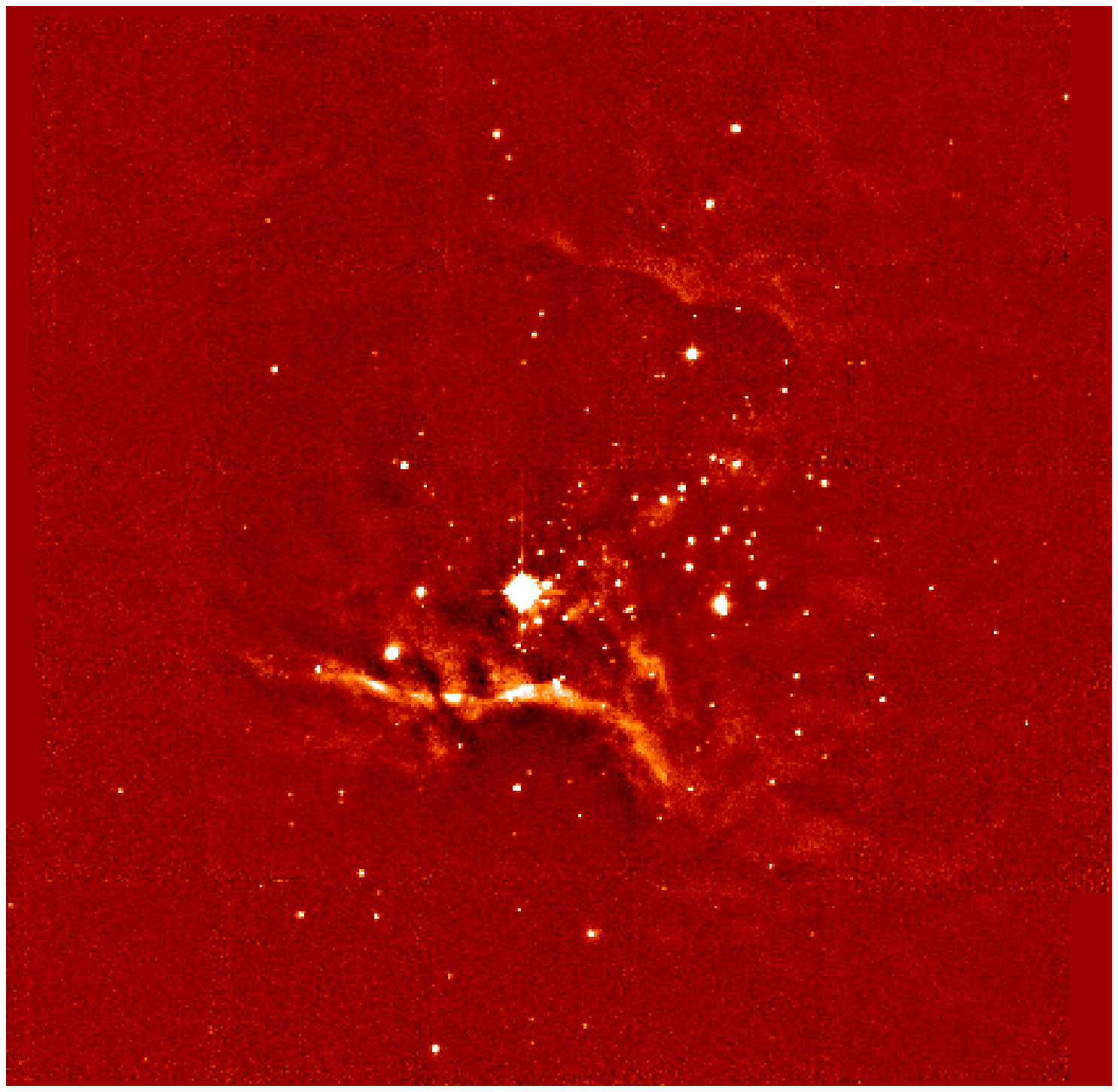]
{An {\it L} band mosaic ({\it top panel}) and a {\it JHK} mosaic ({\it bottom panel}) of the inner 10$\farcm$5 $\times$ 10$\farcm$5 region of the NGC 2024 cluster. The images are displayed with north at the top and west toward the right. The {\it L} mosaic was constructed from 25 individual images each with a total integration time of 3 minutes. The completeness limit for the mosaic is {\it L} = 12.0. The {\it JHK} mosaic was constructed from 9 images each at {\it J,H}, and {\it K} which were combined using the GEOMAP and GEOTRAN packages within IRAF. The completeness limit for the mosaic is {\it K} = 14.0.
\label{N2024_L}
}

\figcaption[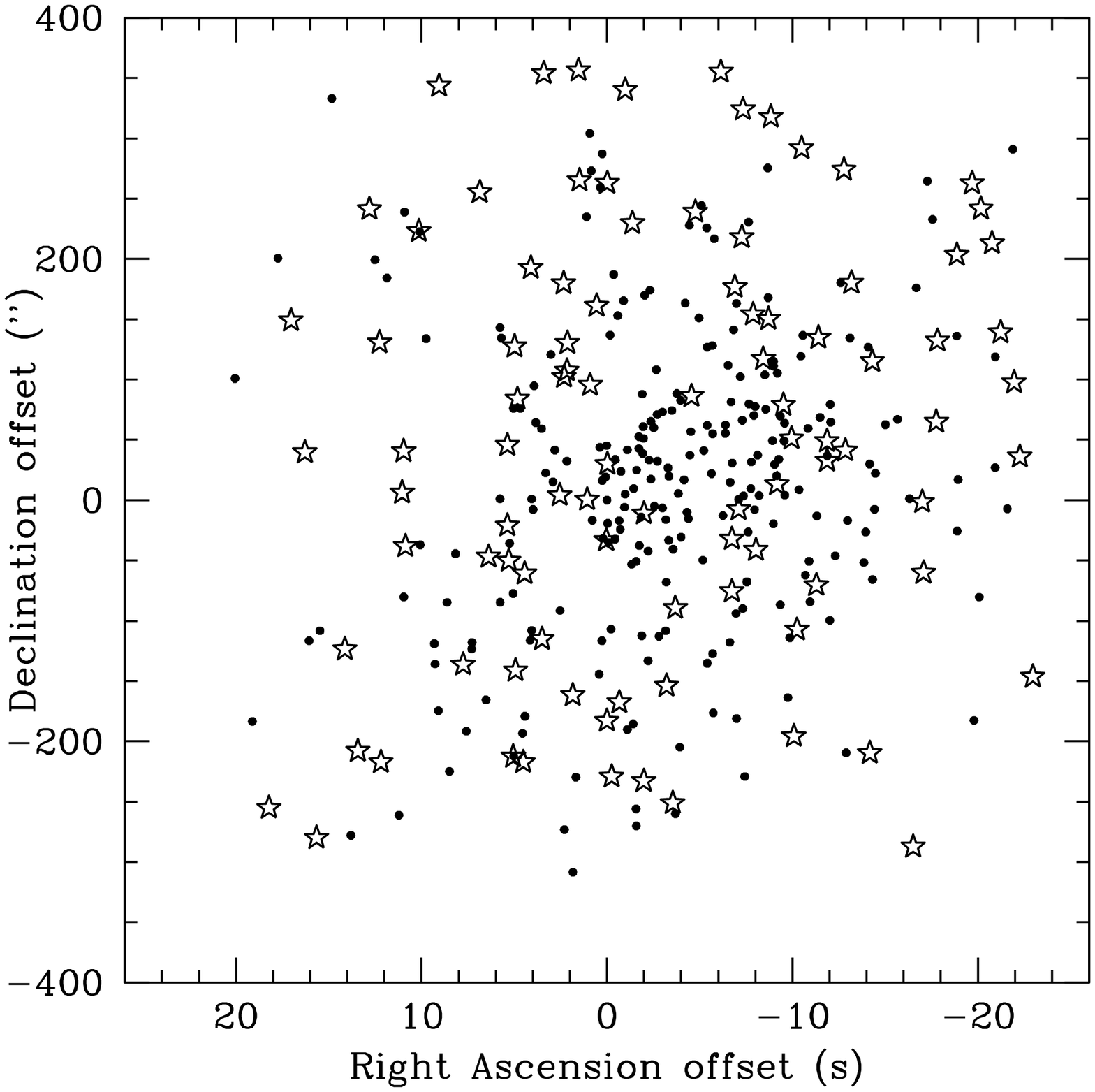]
{
Distribution of {\it K} band sources to the completeness limit of our SQIID survey ({\it K} = 14.5) in NGC 2024. Positions of sources detected
at both {\it K} and {\it L} bands are plotted with a point while sources detected only at {\it K} are shown with a star. The offsets are referred to the position $\alpha$ = 5$^{h}$39$^{m}$13$\stackrel{s}{.}$79, $\delta$ = -1$^{o}$55\arcmin56$\farcs$68 (1950).
\label{clusposn}
}

\figcaption[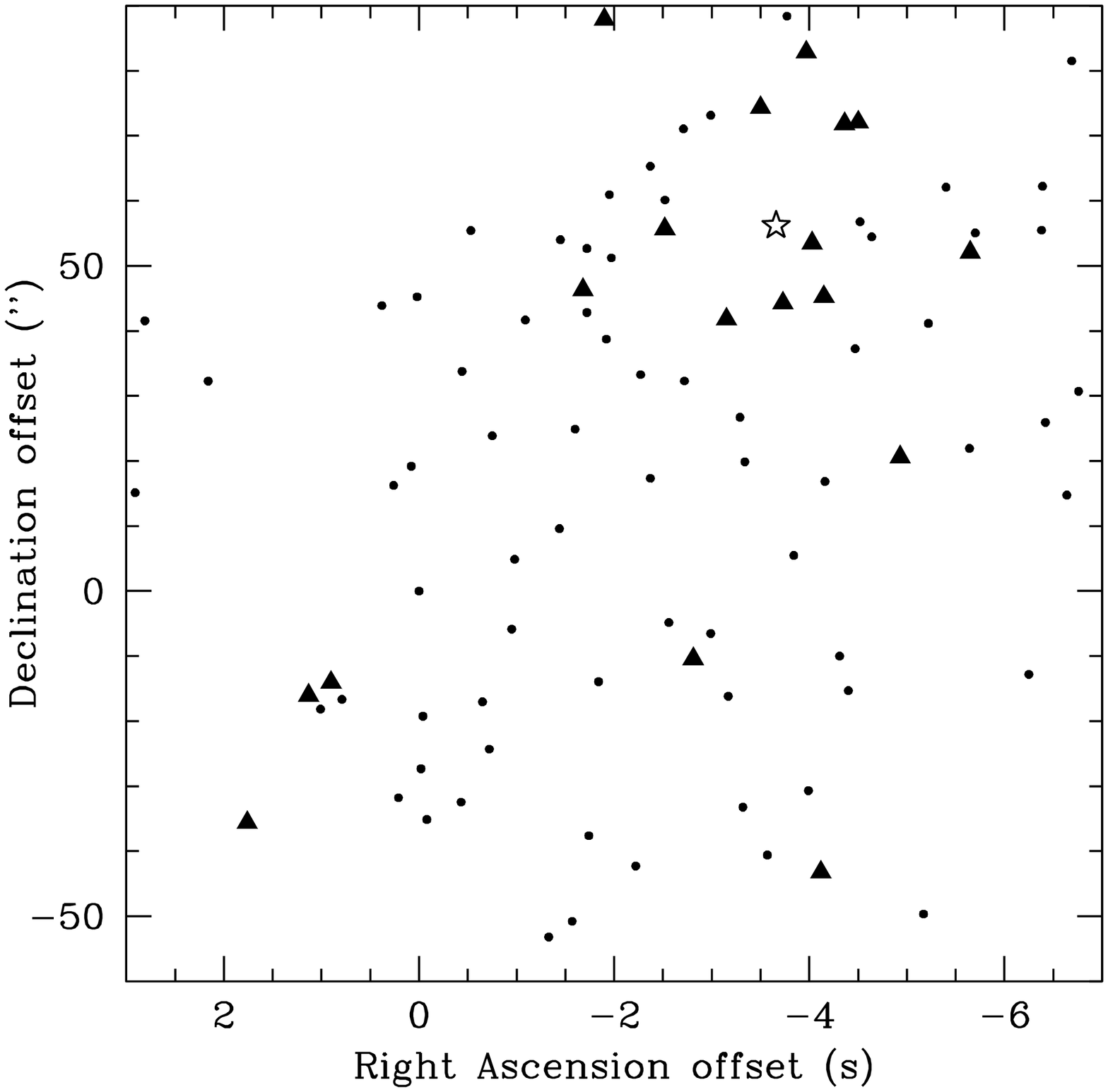]
{
Distribution of sources from our NSFCAM {\it L} band survey. Positions of sources detected which are brighter than our SQIID completeness limit for the central frame ({\it K} =14.0) are plotted with a point while sources having magnitudes fainter than {\it K} = 14.0 are shown with a triangle. A star is used to denote the position of the NSFCAM source detected only at {\it L}. 
\label{cluslonly}
}

\figcaption[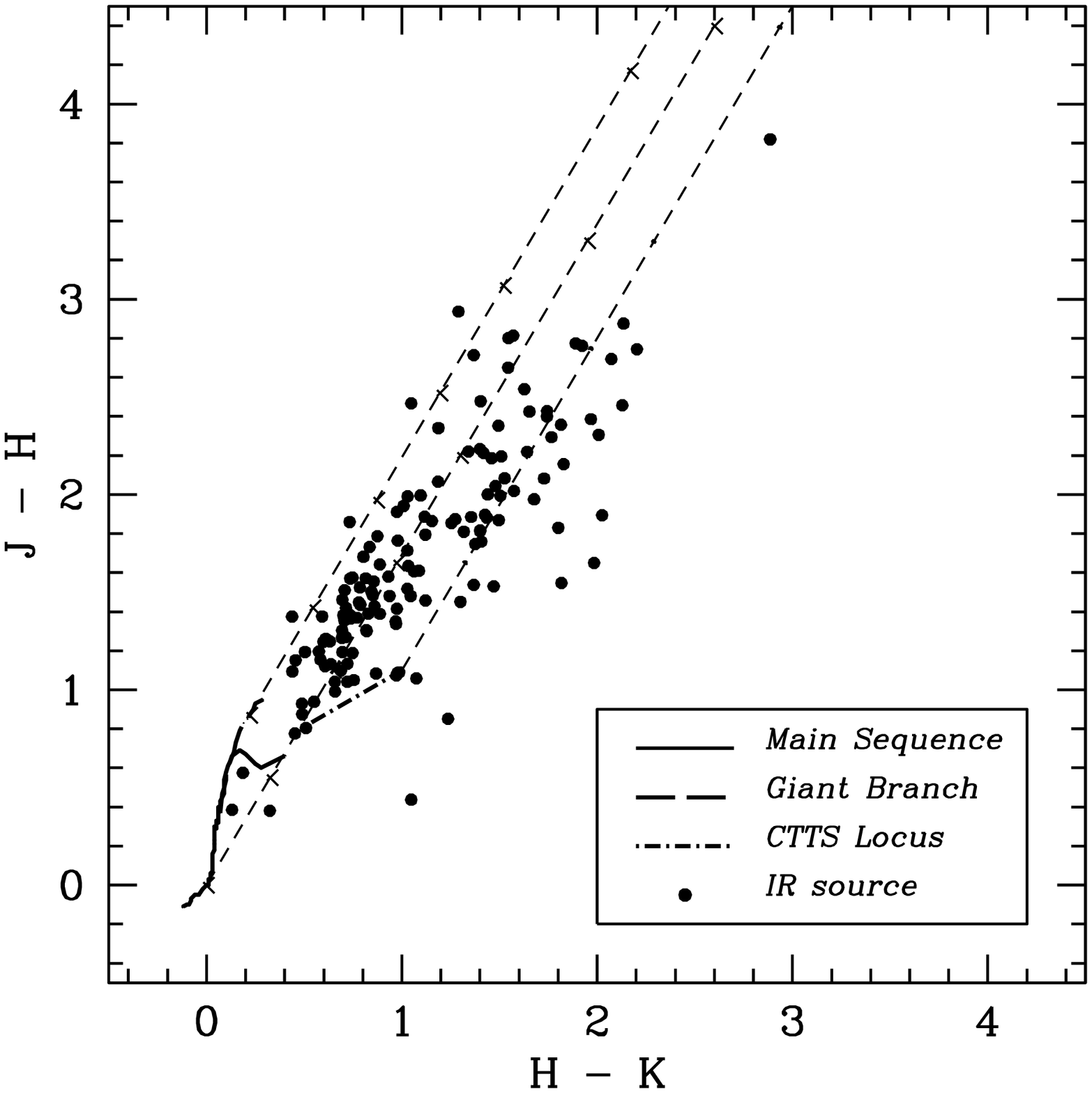]
{
{\it JHK} ({\it left panel}) and {\it JHKL} ({\it right panel}) color-color diagrams for the NGC 2024 cluster. Only those sources extracted from our SQIID survey with {\it K} $\leq$ 12.0 and {\it JHKL} photometric errors less than 10\% are plotted. In the diagram, the locus of points corresponding to the unreddened main sequence is plotted as a solid line, the locus of positions of giant stars is shown as a heavy dashed line and the CTTS locus as a dot-dashed line. The two leftmost dashed lines define the reddening band for main sequence stars and are parallel to the reddening vector. Crosses are placed along these lines at intervals corresponding to 5 magnitudes of visual extinction. The rightmost dashed line is parallel to the reddening band.
\label{ccjhkl}
}

\figcaption[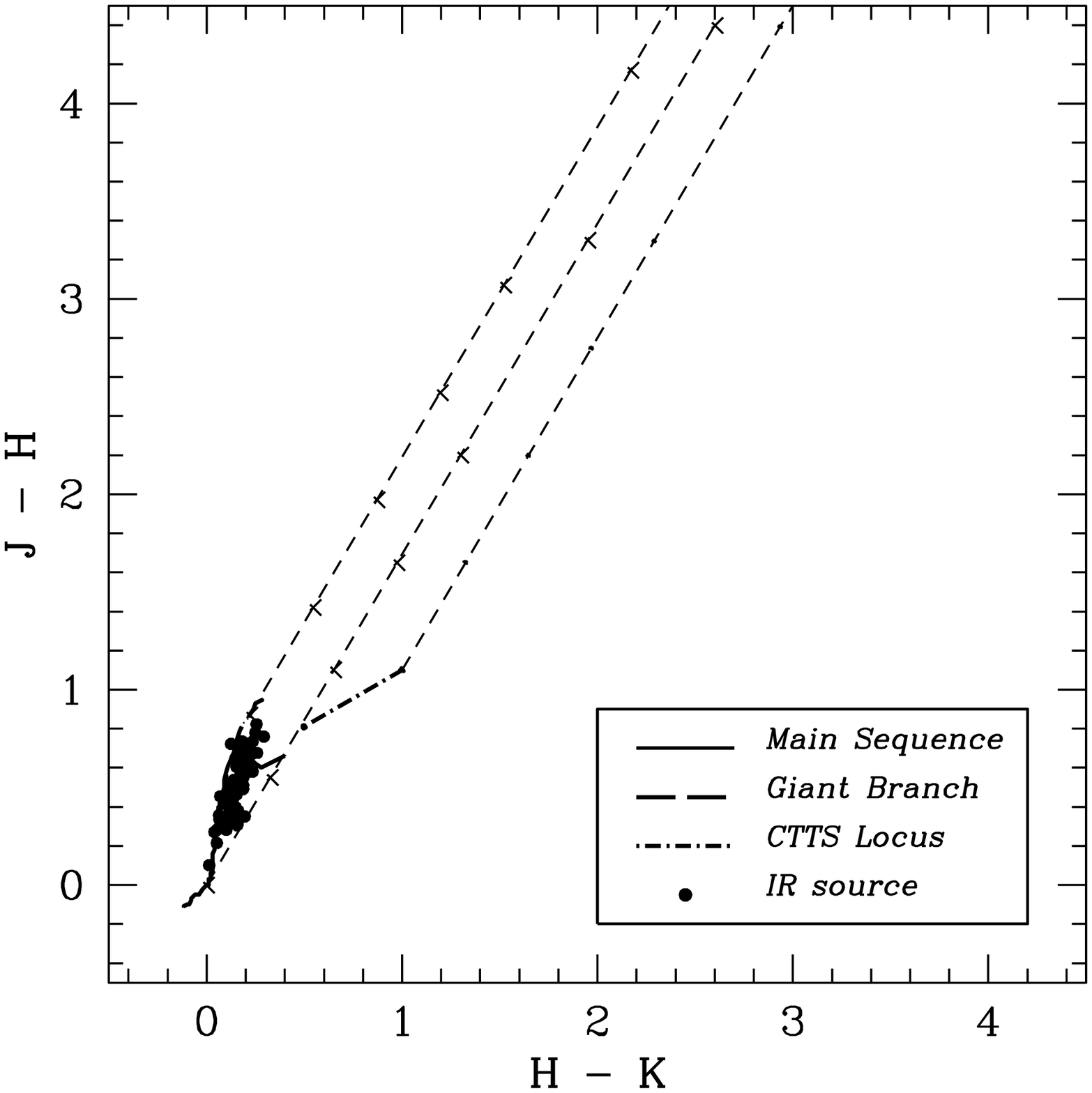]
{
{\it JHK} color-color diagram for the eighteen SQIID control fields. Only those sources extracted from our SQIID and STELIRCAM surveys with {\it K} $\leq$ 12.0 and {\it JHKL} photometric errors less than 10\% are plotted. In the diagram, the locus of points corresponding to the unreddened main sequence is plotted as a solid line, the locus of positions of giant stars is shown as a heavy dashed line and the CTTS locus as a dot-dashed line. The two leftmost dashed lines define the reddening band for main sequence stars and are parallel to the reddening vector. Crosses are placed along these lines at intervals corresponding to 5 magnitudes of visual extinction. The rightmost dashed line is parallel to the reddening band.
\label{ccjhk}
}

\figcaption[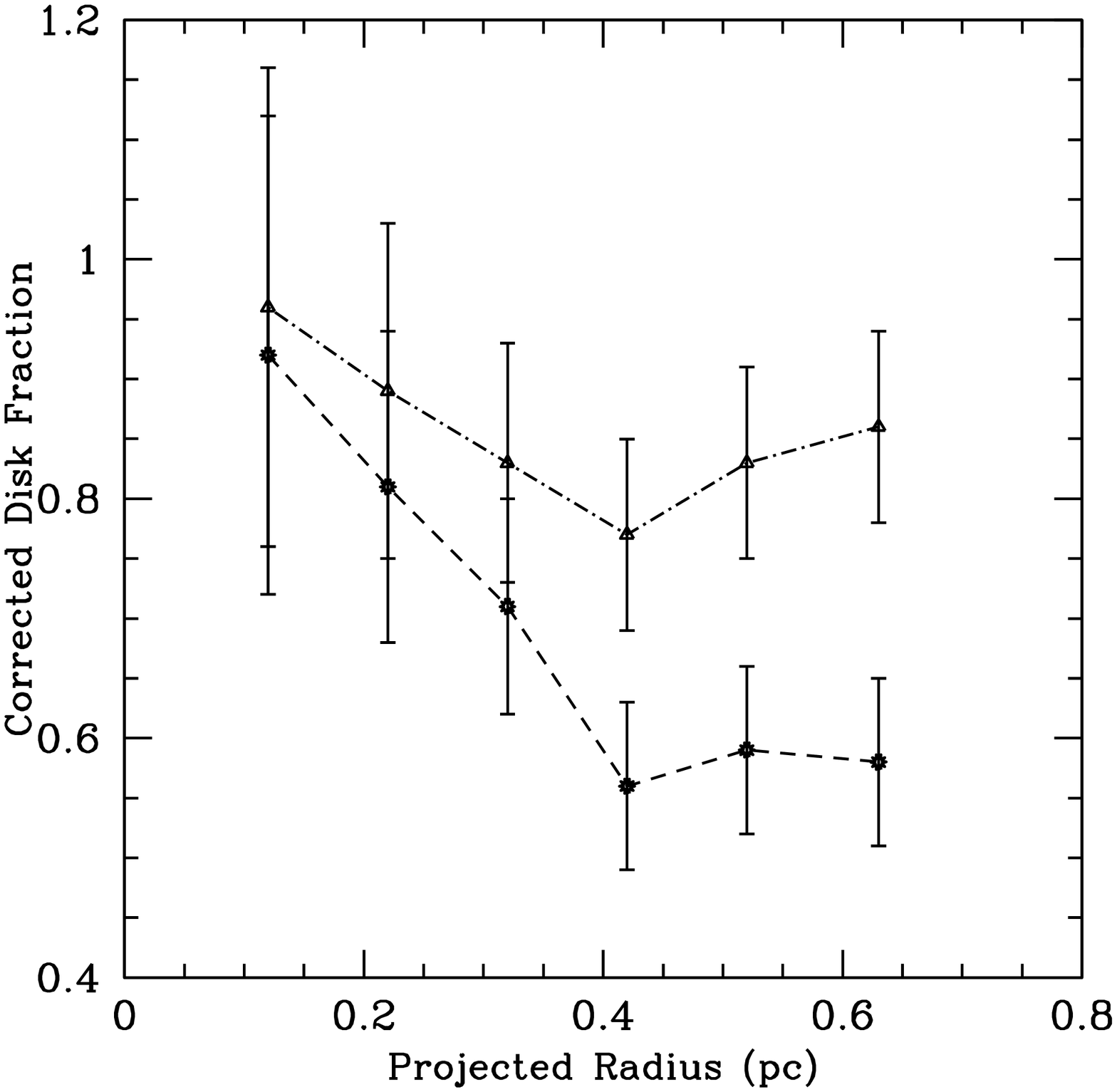]
{
Circumstellar disk fraction vs. projected cluster radius as determined from our {\it K} = {\it L} = 12.0 completeness limit data for the NGC 2024 cluster. The dot-dashed line indicates the disk fraction as determined from {\it JHKL} colors while the dashed line represents the disk fraction determined from {\it JHK} colors. A decrease in the {\it JHKL} circumstellar disk fraction with projected radius is observed, however it is not significant to within the errors. The decrease observed in the {\it JHK} disk fraction {\it is} significant, however. There is a steady decline in the disk fraction to a radius of $\sim$0.4 pc, at which point the circumstellar disk fraction remains constant. The {\it JHK} disk fraction has been corrected for foreground/background star contamination. The error bars represent $\sqrt{{\it N}}$ errors in the disk fractions.
\label{projrad}
}

\figcaption[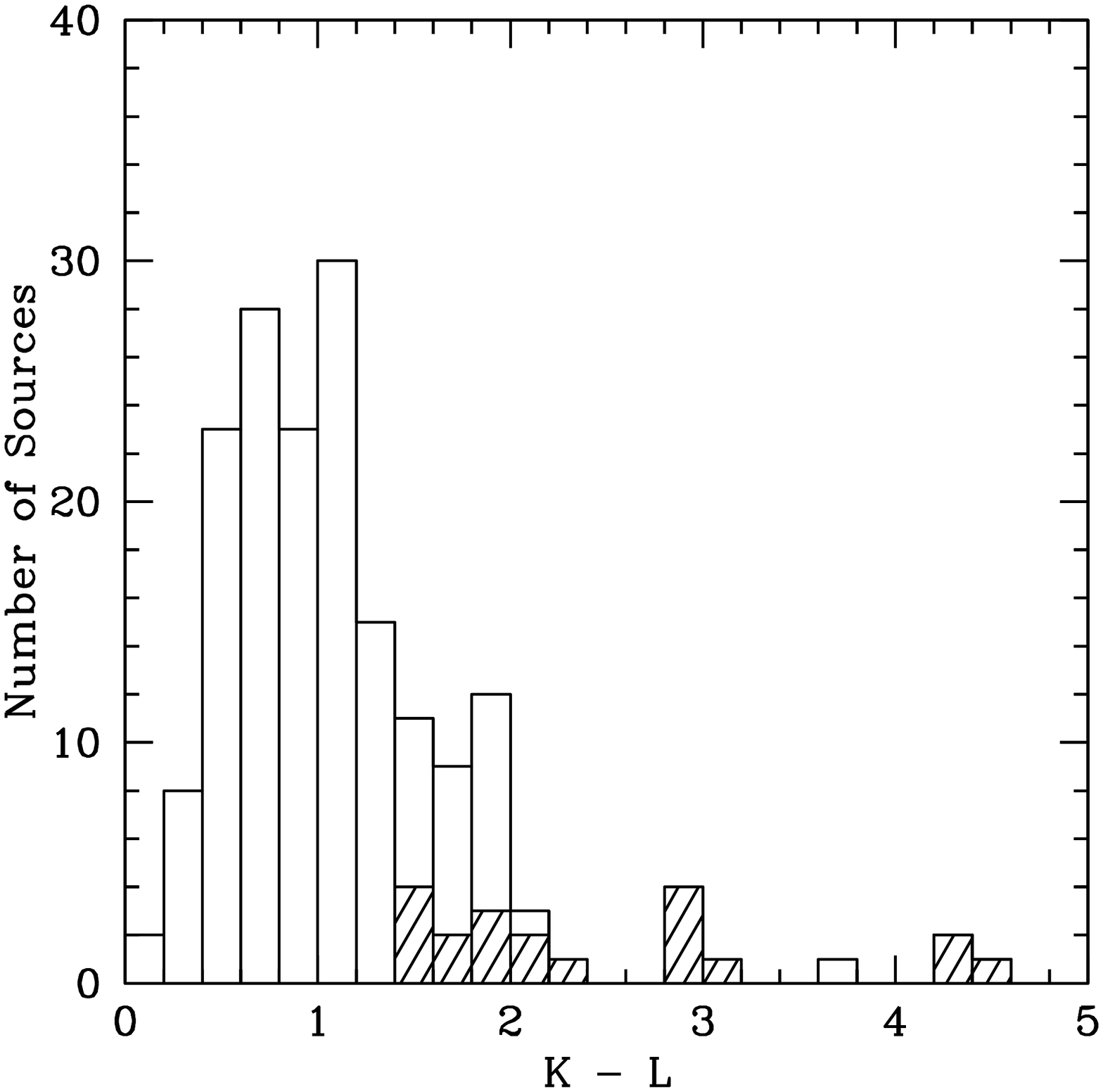]
{
The ({\it K -- L}) frequency distributions of near-infrared colors in the combined STELIRCAM/NSFCAM survey of the NGC 2024 cluster. The shaded region represents those sources from our NSFCAM survey for which {\it K} magnitudes were obtained from Levine et al. 2000 (ie. 14.0 $<$ {\it K} $<$ 16.5). The total number of sources in the distribution is 185.
\label{ngc2024coldist}
}

\figcaption[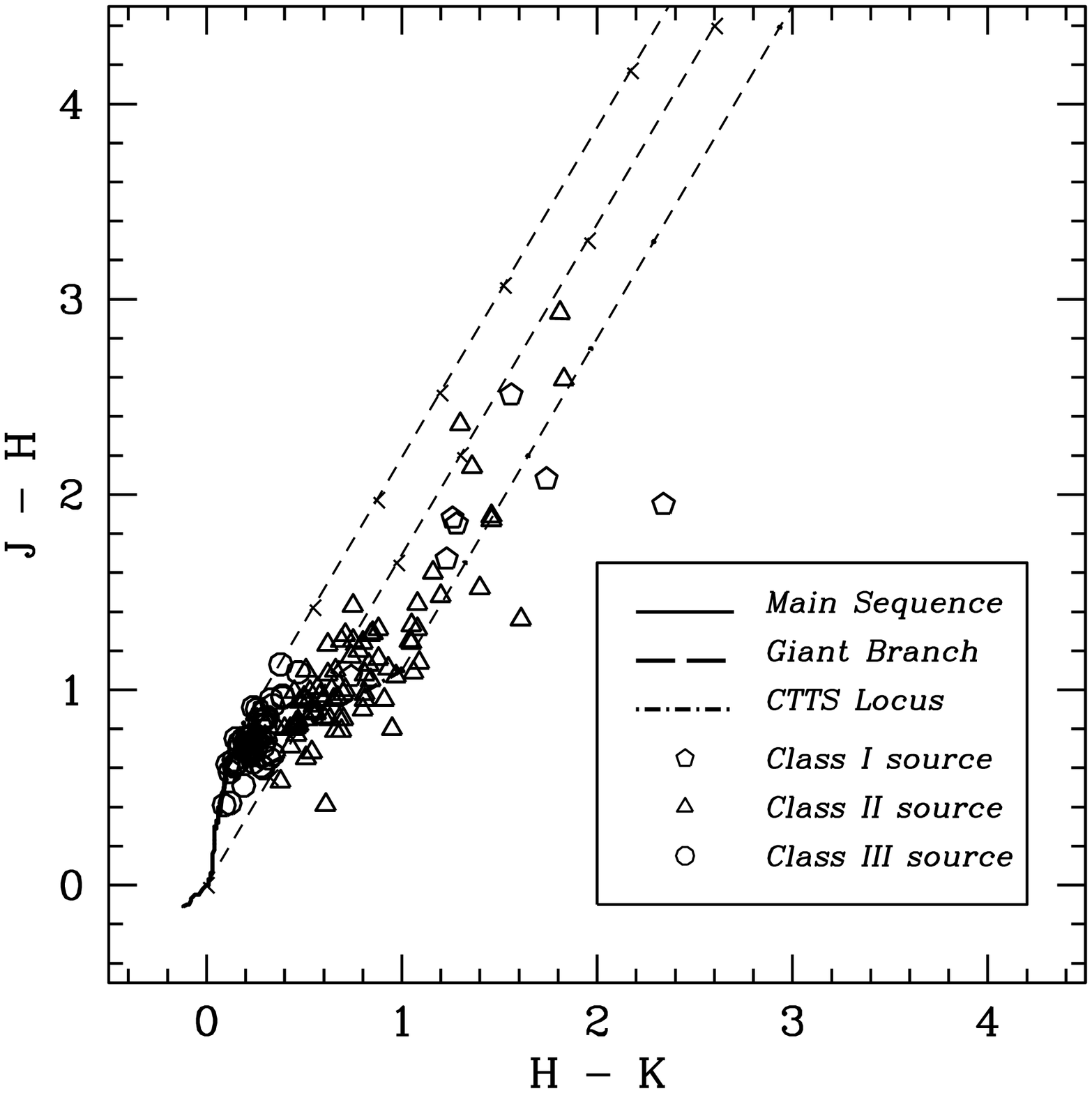]
{
The locations of the Class I, II and III sources in Taurus in the {\it JHK} ({\it left panel}) and {\it JHKL} ({\it right panel}) color-color diagrams. The Class I sources are denoted by a pentagon, Class II sources are shown with a triangle and Class III sources are shown with a circle. The data are taken from Kenyon \& Hartmann (1995).
\label{tauclassjhk}
}

\clearpage

\begin{deluxetable}{cccccc}
\footnotesize
\tablecaption{Summary of Observations \label{sumtable}}
\tablewidth{0pt}
\tablehead{ & & & Integration & Areal & Completeness \nl
Instrument & Telescope & Observation Date & Time (sec) & Coverage (arcmin$^{2}$) & Limits}
\startdata
SQIID   & KPNO 1.3 m &January 1992  & 180 & $\sim$450 & {\it JHK} =
16.5,15.5,14.5 \nl
STELIRCAM & FLWO 1.2 m & January \& December 1998 & 180 & $\sim$110 & {\it L} = 12	 \nl
 & & January 1999 & 1080 & $\sim$20 & {\it L} = 12.9  \nl
NSFCAM & IRTF 3.0 m & January 1999 & 280 & $\sim$6.25 & {\it L} = 14 \nl
\enddata

\end{deluxetable}

\clearpage
\begin{deluxetable}{cccccccccc}
\footnotesize
\tablecaption{{\it JHKL} Photometry of NGC 2024 Sources \label{phottable}}
\tablewidth{0pt}
\tablehead{Star ID & RA (1950)\tablenotemark{1} & Dec (1950)\tablenotemark{1} & J & H & K & L & (J-H) & (H-K) &
(K-L)}
\startdata
   1 &  5  39 15.22 & -1  50 52.42 & 13.16 & 11.80 & 11.10 &  9.88 & 1.36 & 0.70 & 1.22\nl
   2 &  5  38 52.40 & -1  51 05.55 & 10.37 &  9.98 &  9.85 &  9.61 & 0.39 & 0.13 & 0.24\nl
   3 &  5  39 05.61 & -1  51 21.18 & 11.88 &  9.99 &  8.56 &  7.02 & 1.89 & 1.43 & 1.54\nl
   4 &  5  39 15.13 & -1  51 23.46 & 10.96 &  9.66 &  8.84 &  7.70 & 1.30 & 0.82 & 1.14\nl
   5 &  5  39 14.65 & -1  51 37.40 & 12.51 & 11.38 & 10.65 &  9.51 & 1.13 & 0.73 & 1.14\nl
   6 &  5  38 57.00 & -1  51 32.08 & 13.05 & 11.68 & 11.08 & 10.34 & 1.37 & 0.60 & 0.74\nl
   7 &  5  39 25.22 & -1  51 57.71 & 13.88 & 12.50 & 11.77 & 10.81 & 1.38 & 0.73 & 0.96\nl
   8 &  5  39 15.39 & -1  52 01.61 & 12.54 & 11.23 & 10.41 &  9.30 & 1.31 & 0.82 & 1.11\nl
   9 &  5  38 56.72 & -1  52 03.74 & 13.22 & 12.12 & 11.43 & 10.33 & 1.10 & 0.69 & 1.10\nl
  10 &  5  39 06.66 & -1  52 06.09 & 12.25 & 10.39 &  9.23 &  7.99 & 1.86 & 1.16 & 1.24\nl
  11 &  5  39 24.40 & -1  52 14.04 & 14.11 & 12.50 & 11.41 &  9.89 & 1.61 & 1.09 & 1.52\nl
  12 &  5  39 08.50 & -1  52 19.87 & 13.95 & 12.07 & 10.95 &  9.93 & 1.88 & 1.12 & 1.02\nl
  13 &  5  39 13.94 & -1  52 49.55 & 14.72 & 12.73 & 11.71 & 10.61 & 1.99 & 1.02 & 1.10\nl
  14 &  5  39 26.16 & -1  52 52.38 & 13.55 & 12.36 & 11.85 & 11.31 & 1.19 & 0.51 & 0.54\nl
  15 &  5  39 01.68 & -1  52 56.33 & 13.50 & 11.99 & 11.29 & 10.90 & 1.51 & 0.70 & 0.39\nl
  16 &  5  38 57.61 & -1  53 00.59 & 13.03 & 11.94 & 11.50 & 10.97 & 1.09 & 0.44 & 0.53\nl
  17 &  5  39 11.98 & -1  53 02.54 & 15.97 & 13.03 & 11.74 & 10.95 & 2.94 & 1.29 & 0.79\nl
  18 &  5  39 05.59 & -1  53 08.61 & 15.74 & 12.92 & 11.35 &  9.95 & 2.82 & 1.57 & 1.40\nl
  19 &  5  39 13.40 & -1  53 11.24 & 14.21 & 11.73 & 10.33 &  9.43 & 2.48 & 1.40 & 0.90\nl
  20 &  5  39 10.07 & -1  53 13.24 & 13.77 & 12.39 & 11.69 & 11.00 & 1.38 & 0.70 & 0.69\nl
  21 &  5  39 07.31 & -1  53 13.47 & 14.81 & 13.02 & 11.90 & 10.85 & 1.79 & 1.12 & 1.05\nl
  22 &  5  39 20.07 & -1  53 33.61 & 14.70 & 12.89 & 11.49 & 10.12 & 1.81 & 1.40 & 1.37\nl
  23 &  5  39 07.46 & -1  53 35.38 & 11.53 &  9.11 &  7.46 &  5.71 & 2.42 & 1.65 & 1.75\nl
  24 &  5  39 03.72 & -1  53 39.99 & 14.82 & 13.29 & 11.82 &  9.93 & 1.53 & 1.47 & 1.89\nl
  25 &  5  38 55.42 & -1  53 40.48 & 13.78 & 12.58 & 12.01 & 11.29 & 1.20 & 0.57 & 0.72\nl
  26 &  5  39 01.18 & -1  53 42.16 & 13.74 & 12.29 & 11.17 & 10.42 & 1.45 & 1.12 & 0.75\nl
  27 &  5  39 24.05 & -1  53 42.72 & 12.71 & 10.37 &  9.18 &  8.38 & 2.34 & 1.19 & 0.80\nl
  28 &  5  39 08.61 & -1  53 48.43 & 14.94 & 12.59 & 11.09 &  9.89 & 2.35 & 1.42 & 1.20\nl
  29 &  5  39 00.20 & -1  53 49.69 & 12.27 & 10.90 & 10.46 & 10.15 & 1.37 & 0.44 & 0.31\nl
  30 &  5  39 03.84 & -1  53 57.20 & 13.09 & 11.57 & 10.54 &  9.06 & 1.52 & 1.03 & 1.48\nl
  31 &  5  38 53.34 & -1  53 57.69 & 13.64 & 12.28 & 11.54 & 11.01 & 1.36 & 0.74 & 0.53\nl
  32 &  5  39 05.30 & -1  54 01.47 & 12.95 & 11.58 & 10.81 & 10.39 & 1.37 & 0.77 & 0.42\nl
  33 &  5  39 05.09 & -1  54 11.26 & 12.77 & 11.62 & 11.04 & 10.95 & 1.15 & 0.58 & 0.09\nl
  34 &  5  39 05.78 & -1  54 12.70 & 13.19 & 11.75 & 10.96 & 10.09 & 1.44 & 0.79 & 0.87\nl
  35 &  5  39 34.36 & -1  54 15.74 & 12.59 & 11.32 & 10.71 & 10.16 & 1.27 & 0.61 & 0.55\nl
  36 &  5  39 18.23 & -1  54 21.81 & 16.33 & 13.53 & 11.98 & 10.07 & 2.80 & 1.55 & 1.91\nl
  37 &  5  39 06.64 & -1  54 36.80 & 11.95 & 10.18 &  9.20 &  8.51 & 1.77 & 0.98 & 0.69\nl
  38 &  5  39 02.24 & -1  54 37.24 & 13.46 & 12.41 & 11.65 & 10.61 & 1.05 & 0.76 & 1.04\nl
  39 &  5  39 06.30 & -1  54 39.01 & 11.67 & 10.36 &  9.67 &  9.37 & 1.31 & 0.69 & 0.30\nl
  40 &  5  39 19.33 & -1  54 40.46 & 14.87 & 12.40 & 11.35 & 10.25 & 2.47 & 1.05 & 1.10\nl
  41 &  5  39 18.97 & -1  54 40.46 & 09.59 &  8.81 &  8.36 &  8.02 & 0.78 & 0.45 & 0.34\nl
  42 &  5  39 05.73 & -1  54 41.17 & 10.26 &  9.27 &  8.61 &  7.89 & 0.99 & 0.66 & 0.72\nl
  43 &  5  39 04.94 & -1  54 46.93 & 14.08 & 12.66 & 11.69 & 10.95 & 1.42 & 0.97 & 0.74\nl
  44 &  5  39 02.80 & -1  54 47.91 & 12.65 & 11.20 & 10.42 &  9.38 & 1.45 & 0.78 & 1.04\nl
  45 &  5  39 06.99 & -1  54 50.38 & 10.88 &  9.45 &  8.59 &  7.99 & 1.43 & 0.86 & 0.60\nl
  46 &  5  39 11.92 & -1  54 51.29 & 15.38 & 12.98 & 11.24 &  9.72 & 2.40 & 1.74 & 1.52\nl
  47 &  5  39 02.22 & -1  54 51.96 & 11.52 & 10.13 &  9.30 &  8.02 & 1.39 & 0.83 & 1.28\nl
  48 &  5  39 18.13 & -1  54 52.31 & 13.57 & 12.49 & 11.50 & 10.51 & 1.08 & 0.99 & 0.99\nl
  49 &  5  39 07.90 & -1  54 54.37 & 13.19 & 11.00 &  9.54 &  8.15 & 2.19 & 1.46 & 1.39\nl
  50 &  5  39 17.82 & -1  54 57.40 & 13.55 & 12.46 & 11.48 & 10.43 & 1.09 & 0.98 & 1.05\nl
  51 &  5  39 11.77 & -1  54 56.47 & 15.37 & 13.72 & 11.74 & 10.87 & 1.65 & 1.98 & 0.87\nl
  52 &  5  39 09.77 & -1  54 59.81 & 17.72 & 13.90 & 11.01 &  9.02 & 3.82 & 2.89 & 1.99\nl
  53 &  5  39 08.59 & -1  55 01.54 & 14.11 & 11.72 &  9.76 &  7.90 & 2.39 & 1.96 & 1.86\nl
  54 &  5  39 07.91 & -1  55 01.13 & 14.03 & 12.49 & 11.12 &  9.73 & 1.54 & 1.37 & 1.39\nl
  55 &  5  39 05.36 & -1  55 07.33 & 14.82 & 12.75 & 11.57 & 10.41 & 2.07 & 1.18 & 1.16\nl
  56 &  5  39 04.74 & -1  55 07.47 & 14.14 & 12.35 & 11.48 & 10.54 & 1.79 & 0.87 & 0.94\nl
  57 &  5  39 14.31 & -1  55 11.35 & 14.74 & 12.98 & 11.58 & 10.35 & 1.76 & 1.40 & 1.23\nl
  58 &  5  39 14.68 & -1  55 12.72 & 13.68 & 11.46 & 10.12 &  9.19 & 2.22 & 1.34 & 0.93\nl
  59 &  5  39 17.10 & -1  55 15.05 & 14.68 & 12.82 & 11.57 & 10.60 & 1.86 & 1.25 & 0.97\nl
  60 &  5  39 09.82 & -1  55 19.34 & 16.64 & 13.88 & 11.96 & 10.26 & 2.76 & 1.92 & 1.70\nl
  61 &  5  39 06.17 & -1  55 19.26 & 10.71 &  9.16 &  8.30 &  7.70 & 1.55 & 0.86 & 0.60\nl
  62 &  5  39 02.40 & -1  55 19.81 & 13.84 & 12.11 & 11.27 & 10.03 & 1.73 & 0.84 & 1.24\nl
  63 &  5  39 05.03 & -1  55 22.70 & 11.93 & 10.40 &  9.62 &  9.17 & 1.53 & 0.78 & 0.45\nl
  64 &  5  39 06.51 & -1  55 24.95 & 13.46 & 11.51 & 10.50 &  9.83 & 1.95 & 1.01 & 0.67\nl
  65 &  5  39 07.53 & -1  55 25.91 & 12.17 & 10.29 &  8.93 &  7.88 & 1.88 & 1.36 & 1.05\nl
  66 &  5  39 00.12 & -1  55 26.61 & 12.43 & 11.18 & 10.55 &  9.89 & 1.25 & 0.63 & 0.66\nl
  67 &  5  39 05.25 & -1  55 27.01 & 12.07 & 10.59 &  9.65 &  8.88 & 1.48 & 0.94 & 0.77\nl
  68 &  5  39 11.00 & -1  55 29.88 & 15.10 & 13.20 & 11.18 &  9.85 & 1.90 & 1.25 & 1.33\nl
  69 &  5  39 17.60 & -1  55 34.14 & 13.99 & 12.92 & 11.95 & 10.92 & 1.07 & 0.97 & 1.03\nl
  70 &  5  39 10.95 & -1  55 36.74 & 13.87 & 12.13 & 10.75 & 10.02 & 1.74 & 1.38 & 0.73\nl
  71 &  5  39 05.13 & -1  55 36.51 & 13.33 & 11.76 & 10.95 & 10.37 & 1.57 & 0.81 & 0.58\nl
  72 &  5  38 55.35 & -1  55 39.51 & 13.48 & 12.39 & 11.53 & 10.36 & 1.09 & 0.86 & 1.17\nl
  73 &  5  39 10.13 & -1  55 39.76 & 15.41 & 13.11 & 11.10 &  9.45 & 2.30 & 2.01 & 1.65\nl
  74 &  5  39 17.20 & -1  55 41.48 & 15.39 & 13.35 & 11.87 & 10.14 & 2.04 & 1.48 & 1.73\nl
  75 &  5  39 07.64 & -1  55 41.84 & 13.60 & 11.24 &  9.43 &  7.44 & 2.36 & 1.81 & 1.99\nl
  76 &  5  39 12.85 & -1  55 47.01 & 14.58 & 12.50 & 10.77 &  9.60 & 2.08 & 1.73 & 1.17\nl
  77 &  5  39 13.31 & -1  55 51.72 & 13.02 & 10.59 &  8.85 &  7.11 & 2.43 & 1.74 & 1.74\nl
  78 &  5  39 10.45 & -1  55 51.13 & 14.32 & 12.03 & 10.26 &  8.40 & 2.29 & 1.77 & 1.86\nl
  79 &  5  39 06.93 & -1  55 53.08 & 11.74 & 11.36 & 11.03 & 10.98 & 0.38 & 0.33 & 0.05\nl
  80 &  5  39 06.09 & -1  55 52.75 & 12.51 & 11.12 & 10.24 &  9.70 & 1.39 & 0.88 & 0.54\nl
  81 &  5  39 04.70 & -1  55 52.45 & 12.38 & 10.51 &  9.01 &  7.20 & 1.87 & 1.50 & 1.81\nl
  82 &  5  38 57.96 & -1  55 55.43 & 11.73 & 10.61 & 10.00 &  9.30 & 1.12 & 0.61 & 0.74\nl
  84 &  5  39 13.34 & -1  56 02.46 & 14.18 & 12.02 & 10.20 &  8.64 & 2.16 & 1.82 & 1.56\nl
  85 &  5  39 11.30 & -1  56 03.13 & 14.08 & 11.87 & 10.45 &  9.52 & 2.21 & 1.42 & 0.93\nl
  86 &  5  39 06.33 & -1  56 04.40 &  7.99 &  7.55 &  6.50 &  5.60 & 0.44 & 1.05 & 0.90\nl
  87 &  5  39 09.98 & -1  56 06.61 & 15.83 & 12.96 & 10.82 &  9.12 & 2.87 & 2.14 & 1.70\nl
  88 &  5  39 02.97 & -1  56 09.70 & 11.72 & 10.59 &  9.96 &  9.55 & 1.13 & 0.63 & 0.41\nl
  89 &  5  39 13.64 & -1  56 13.63 & 14.32 & 11.57 &  9.37 &  7.46 & 2.75 & 2.20 & 1.91\nl
  90 &  5  39 11.12 & -1  56 12.79 & 14.97 & 12.44 & 10.81 &  9.92 & 2.53 & 1.63 & 0.89\nl
  91 &  5  39 14.25 & -1  56 15.85 & 13.79 & 11.96 & 10.16 &  9.05 & 1.83 & 1.80 & 1.11\nl
  92 &  5  39 05.32 & -1  56 16.24 & 13.75 & 12.41 & 11.44 & 10.70 & 1.34 & 0.97 & 0.74\nl
  93 &  5  38 55.39 & -1  56 22.27 & 12.44 & 11.51 & 11.02 & 10.35 & 0.93 & 0.49 & 0.67\nl
  94 &  5  39 06.67 & -1  56 23.03 & 13.19 & 12.13 & 11.06 & 10.29 & 1.06 & 1.07 & 0.77\nl
  95 &  5  39 00.33 & -1  56 23.14 & 13.39 & 11.81 & 11.07 & 10.62 & 1.58 & 0.74 & 0.45\nl
  96 &  5  39 14.50 & -1  56 28.38 & 15.23 & 12.53 & 10.46 &  8.68 & 2.70 & 2.07 & 1.78\nl
  97 &  5  39 10.97 & -1  56 29.83 & 15.64 & 13.19 & 11.06 &  8.96 & 2.45 & 2.13 & 2.10\nl
  98 &  5  39 19.54 & -1  56 32.45 & 12.21 & 10.40 &  9.09 &  7.20 & 1.81 & 1.31 & 1.89\nl
  99 &  5  39 24.37 & -1  56 33.61 & 12.99 & 11.95 & 11.23 & 10.91 & 1.04 & 0.72 & 0.32\nl
 100 &  5  39 10.72 & -1  56 37.21 & 12.67 & 11.41 & 10.71 & 10.02 & 1.26 & 0.70 & 0.69\nl
 101 &  5  39 22.47 & -1  56 41.00 & 14.03 & 11.95 & 10.42 &  8.44 & 2.08 & 1.53 & 1.98\nl
 102 &  5  39 01.97 & -1  56 42.76 & 14.10 & 12.64 & 11.94 & 11.13 & 1.46 & 0.70 & 0.81\nl
 103 &  5  39 09.12 & -1  56 46.26 & 15.00 & 12.78 & 11.14 &  9.73 & 2.22 & 1.64 & 1.41\nl
 104 &  5  39 03.39 & -1  56 47.23 & 10.75 &  9.82 &  9.27 &  8.81 & 0.93 & 0.55 & 0.46\nl
 105 &  5  39 12.72 & -1  56 47.38 & 14.26 & 12.28 & 10.60 &  9.04 & 1.98 & 1.68 & 1.56\nl
 106 &  5  39 00.43 & -1  56 48.34 & 12.30 & 10.72 &  9.79 &  8.85 & 1.58 & 0.93 & 0.94\nl
 107 &  5  39 12.96 & -1  56 49.79 & 13.19 & 11.17 &  9.60 &  8.19 & 2.02 & 1.57 & 1.41\nl
 108 &  5  38 59.97 & -1  57 02.42 & 13.23 & 11.78 & 10.48 &  9.32 & 1.45 & 1.30 & 1.16\nl
 109 &  5  39 06.75 & -1  57 04.36 & 13.08 & 11.89 & 11.15 & 10.24 & 1.19 & 0.74 & 0.91\nl
 110 &  5  39 25.26 & -1  57 16.88 & 13.26 & 12.38 & 11.89 & 11.44 & 0.88 & 0.49 & 0.45\nl
 111 &  5  38 54.21 & -1  57 17.00 & 13.87 & 12.26 & 11.20 &  9.89 & 1.61 & 1.06 & 1.31\nl
 112 &  5  39 22.92 & -1  57 21.39 & 13.19 & 12.15 & 11.49 & 11.05 & 1.04 & 0.66 & 0.44\nl
 113 &  5  39 04.95 & -1  57 23.31 & 12.78 & 11.53 & 10.93 & 10.49 & 1.25 & 0.60 & 0.44\nl
 114 &  5  39 07.33 & -1  57 30.55 & 14.34 & 12.35 & 11.25 & 10.22 & 1.99 & 1.10 & 1.03\nl
 115 &  5  39 02.26 & -1  57 36.24 & 14.96 & 12.77 & 11.26 & 10.11 & 2.19 & 1.51 & 1.15\nl
 116 &  5  39 14.07 & -1  57 43.65 & 11.73 & 11.16 & 10.97 & 10.56 & 0.57 & 0.19 & 0.41\nl
 117 &  5  39 11.13 & -1  57 44.99 & 13.63 & 12.15 & 11.30 & 10.81 & 1.48 & 0.85 & 0.49\nl
 118 &  5  39 04.41 & -1  57 50.66 & 14.70 & 12.71 & 11.20 &  9.78 & 1.99 & 1.51 & 1.42\nl
 119 &  5  39 18.45 & -1  57 52.96 & 13.95 & 12.60 & 11.63 & 11.34 & 1.35 & 0.97 & 0.29\nl
 120 &  5  39 30.36 & -1  57 53.18 & 13.14 & 11.66 & 10.61 &  9.62 & 1.48 & 1.05 & 0.99\nl
 121 &  5  39 21.57 & -1  57 54.61 & 15.18 & 13.64 & 11.82 &  9.95 & 1.54 & 1.82 & 1.87\nl
 122 &  5  39 07.66 & -1  57 54.51 & 14.43 & 12.55 & 11.11 &  9.97 & 1.88 & 1.44 & 1.14\nl
 123 &  5  39 21.59 & -1  57 59.99 & 13.52 & 11.88 & 10.99 & 10.36 & 1.64 & 0.89 & 0.63\nl
 124 &  5  39 12.07 & -1  58 09.75 & 15.31 & 12.60 & 11.23 & 10.48 & 2.71 & 1.37 & 0.75\nl
 125 &  5  39 23.57 & -1  58 12.49 & 14.14 & 12.46 & 11.65 & 10.57 & 1.68 & 0.81 & 1.08\nl
 126 &  5  39 04.54 & -1  58 40.33 & 15.27 & 12.62 & 11.08 & 10.20 & 2.65 & 1.54 & 0.88\nl
 127 &  5  39 20.83 & -1  58 42.20 & 12.87 & 11.23 & 10.20 &  9.03 & 1.64 & 1.03 & 1.17\nl
 128 &  5  39 23.39 & -1  58 51.22 & 12.88 & 11.73 & 11.28 & 10.75 & 1.15 & 0.45 & 0.53\nl
 129 &  5  39 18.72 & -1  58 55.88 & 13.41 & 11.99 & 11.28 & 10.67 & 1.42 & 0.71 & 0.61\nl
 130 &  5  39 07.30 & -1  58 57.71 & 13.59 & 12.32 & 11.61 & 10.79 & 1.27 & 0.71 & 0.82\nl
 131 &  5  38 54.49 & -1  58 59.23 & 13.61 & 12.10 & 11.26 & 10.65 & 1.51 & 0.84 & 0.61\nl
 132 &  5  39 33.42 & -1  59 00.09 & 13.77 & 11.91 & 11.18 & 10.44 & 1.86 & 0.73 & 0.74\nl
 133 &  5  39 13.20 & -1  59 06.81 & 13.67 & 11.76 & 10.79 & 10.15 & 1.91 & 0.97 & 0.64\nl
 134 &  5  39 21.88 & -1  59 08.14 & 13.46 & 11.65 & 10.25 &  8.55 & 1.81 & 1.40 & 1.70\nl
 135 &  5  39 18.84 & -1  59 10.00 & 12.62 & 11.77 & 10.53 &  9.36 & 0.85 & 1.24 & 1.17\nl
 136 &  5  39 10.36 & -1  59 21.45 & 12.30 & 10.43 &  9.15 &  7.93 & 1.87 & 1.28 & 1.22\nl
 137 &  5  39 06.85 & -1  59 45.77 & 14.16 & 12.59 & 11.86 & 11.01 & 1.57 & 0.73 & 0.85\nl
 138 &  5  39 15.97 & -1  59 46.31 & 16.55 & 13.77 & 11.88 &  9.77 & 2.78 & 1.89 & 2.11\nl
 139 &  5  39 12.72 & -2  00 12.63 & 13.48 & 12.29 & 11.59 & 10.26 & 1.19 & 0.70 & 1.33\nl
 140 &  5  39 16.58 & -2  00 29.75 & 12.89 & 10.89 &  9.45 &  7.43 & 2.00 & 1.44 & 2.02\nl
 141 &  5  39 28.10 & -2  00 34.59 & 12.94 & 12.13 & 11.62 & 10.91 & 0.81 & 0.51 & 0.71\nl
 142 &  5  39 16.13 & -2  01 05.13 & 15.16 & 12.93 & 11.53 & 10.69 & 2.23 & 1.40 & 0.84\nl
\enddata
\tablenotetext{1}{Coordinates are approximate, and accurate to only a few
arcseconds. See text for description of how coordinates were derived.}
\end{deluxetable}

\clearpage

\begin{deluxetable}{ccccc}
\small
\tablecaption{Infrared Excess Fractions for the NGC 2024 Cluster\tablenotemark{1}\label{fractable}}
\tablewidth{0pt}
\tablehead{
\hspace*{0.5in} &\colhead{Area (arcmin$^{2}$)\tablenotemark{2}} & &
\colhead{{\it JHK} Fraction} & \colhead{{\it JHKL} Fraction}
}
\startdata
\cutinhead{\hspace*{1.5in}a. Completeness Limit {\it K} = {\it L} =
12.0\tablenotemark{3}}
 & 110.25 &  & 76/131 (58$\pm$7\%) & 112/131 (86$\pm$8\%) \nl
 & 20.25 &  & 58/82 (71$\pm$9\%) & 68/82 (83$\pm$10\%) \nl
 & 6.25 &  & 28/32 (88$\pm$17\%) & 30/32 (94$\pm$17\%) \nl
\tablevspace{0.1in}
\cutinhead{\hspace*{1.5in}b. Completeness Limit {\it K} = {\it L} =
12.9\tablenotemark{4}}
 & 20.25 &  & 58/90 (64$\pm$9\%) & 76/90 (84$\pm$10\%) \nl
 & 6.25 &  & 33/38 (87$\pm$15\%) & 36/38 (95$\pm$16\%) \nl
\tablevspace{0.1in}
\cutinhead{\hspace*{1.5in}c. Completeness Limit {\it K} = {\it L} =
14.0\tablenotemark{5}}
 & 6.25 &  & 34/38 (90$\pm$15\%) & 33/38 (87$\pm$15\%) \nl
\enddata

\tablenotetext{1}{Infrared excess fractions have been corrected for field star contamination.}
\tablenotetext{2}{Area centered on the cluster for which infrared excess fractions were
determined.}
\tablenotetext{3}{Completeness limit for 3 minute SAO data.}
\tablenotetext{4}{Completeness limit for 18 minute SAO data.}
\tablenotetext{5}{Completeness limit for IRTF data.}

\end{deluxetable}

\clearpage

\clearpage
\plotfiddle{Haisch.fig1a.ps}{3.0in}{0.0}{65}{67}{-200}{-200}
\plotfiddle{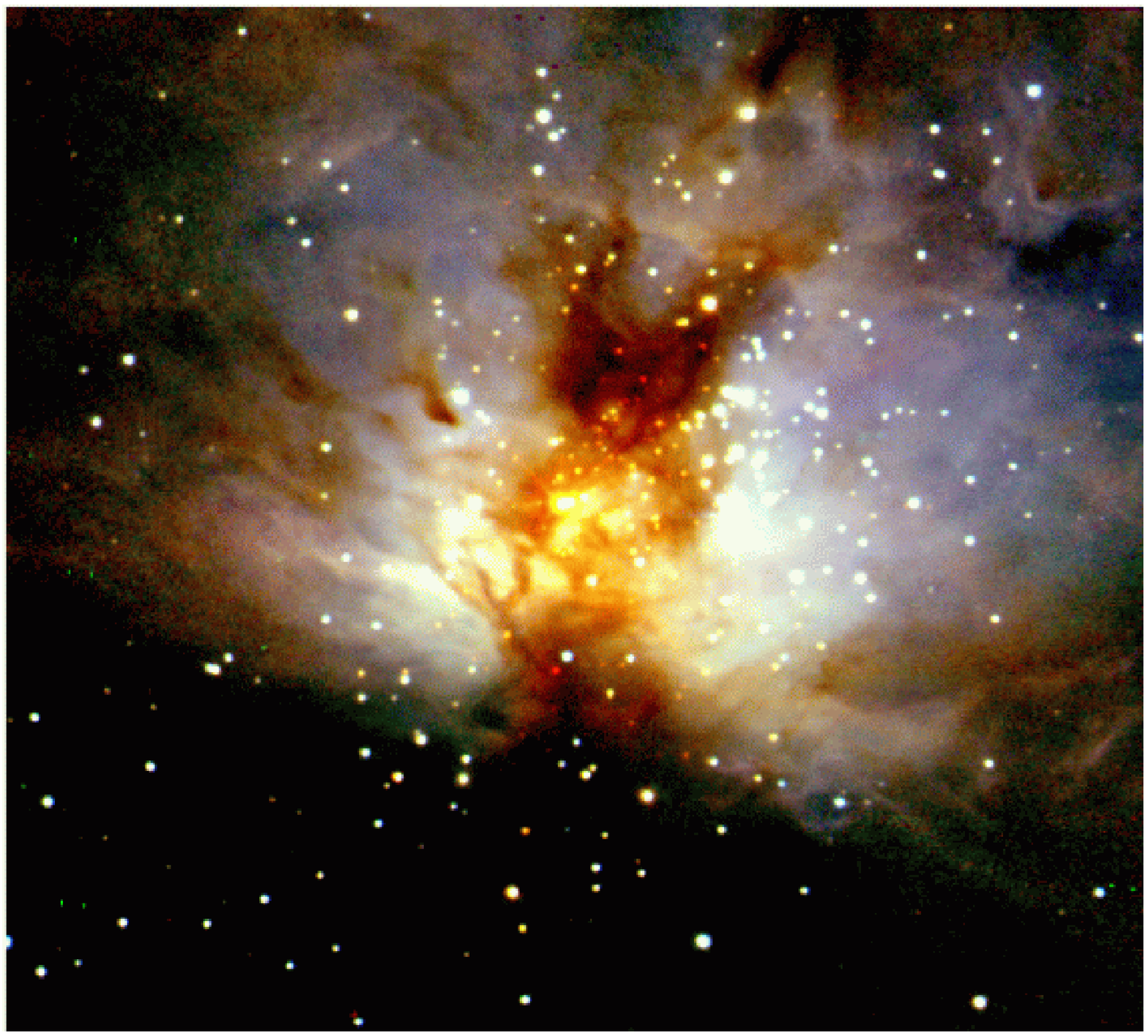}{1.0in}{0.0}{48.5}{51}{-150}{-330}
\clearpage
\plotone{Haisch.fig2.eps}
\clearpage
\plotone{Haisch.fig3.eps}
\clearpage
\plotone{Haisch.fig4a.eps}
\clearpage
\plotone{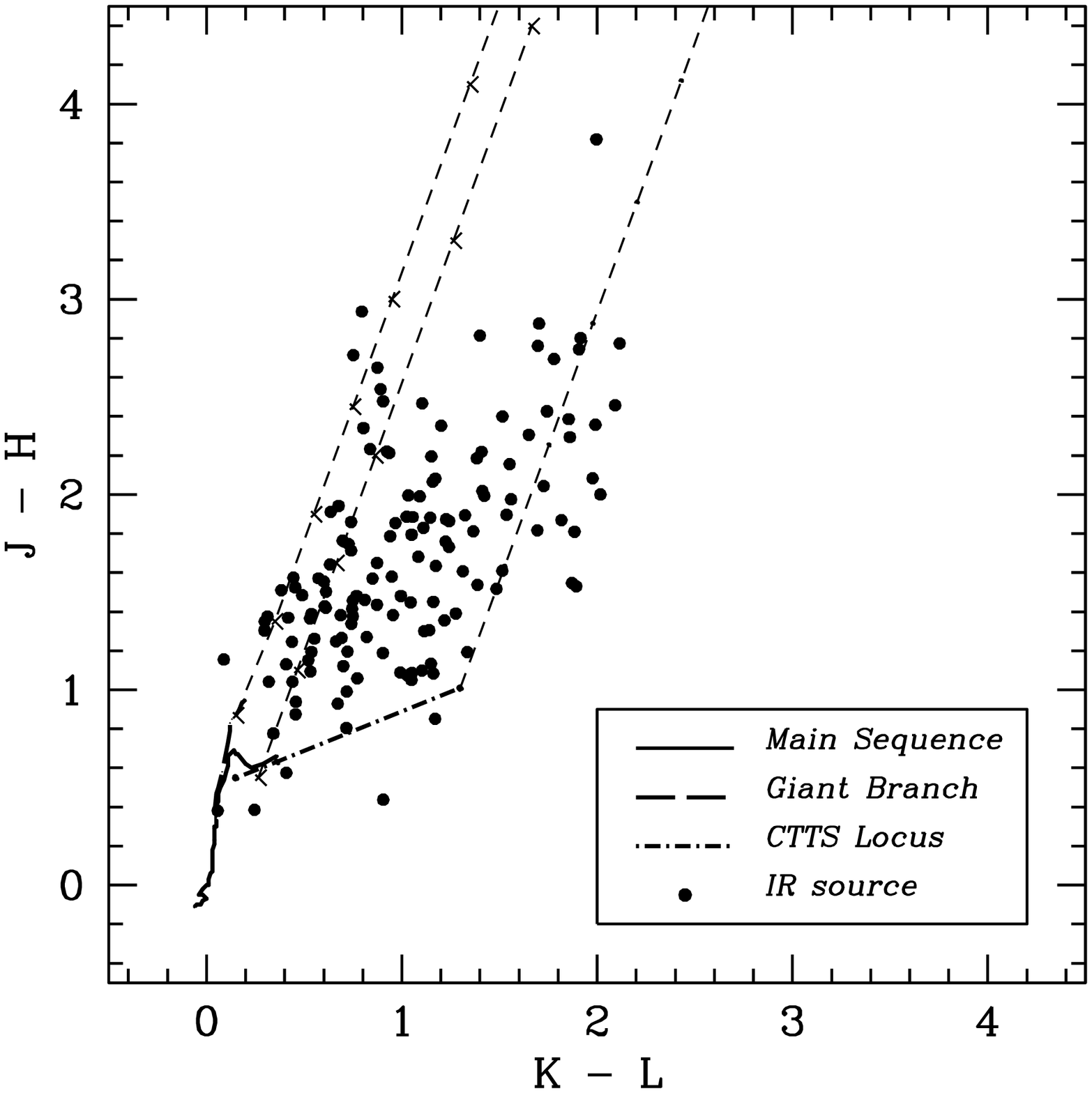}
\clearpage
\plotone{Haisch.fig5.eps}
\clearpage
\plotone{Haisch.fig6.eps}
\clearpage
\plotone{Haisch.fig7.eps}
\clearpage
\plotone{Haisch.fig8a.eps}
\clearpage
\plotone{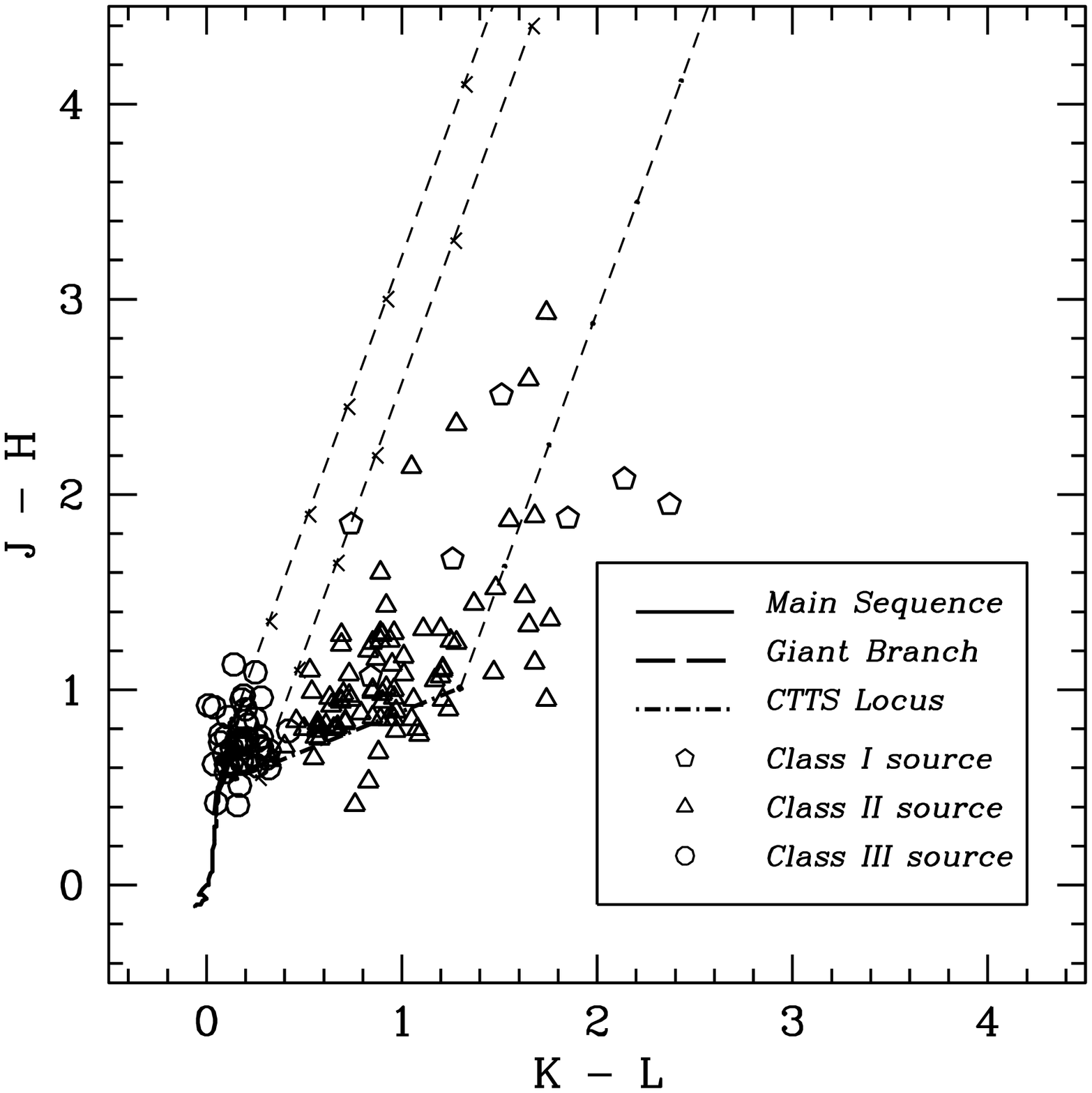}
\clearpage

\end{document}